\documentclass[10pt,journal]{IEEEtran}
\usepackage{amssymb,amsfonts}
\usepackage{amsmath,algorithm,algpseudocode,color,bm,theorem}
\usepackage{tabularx}
\usepackage{epsfig}
\usepackage{graphicx}
\usepackage[T1]{fontenc}
\usepackage{epstopdf}
\usepackage{cite,subfigure,cuted}
\usepackage{rotating,setspace,latexsym,dsfont,caption}
\usepackage{lipsum}

\newtheorem{Theorem}{Theorem}
\newtheorem{Corollary}{Corollary}
\newtheorem{Lemma}{Lemma}

\newtheorem{Definition}{Definition}

\newenvironment{Proof}[1]{\medskip\par\noindent
	{\bf Proof:\,}\,#1}{{\mbox{\,$\blacksquare$}\par}}

\newcommand{\xv}{\mathbf{x}}

\newcommand{\sv}{\mathbf{s}}

\newcommand{\Sv}{\mathbf{S}}

\newcommand{\Eb}{\mathbb{E}}
\newcommand{\Pb}{\mathbb{P}}
\newcommand{\Zb}{\mathbb{Z}}

\newcommand{\lv}{\mathbf{1}}

\newcommand{\Sc}{\mathcal{S}}

\newcommand{\Fc}{\mathcal{F}}

\newcommand{\Hc}{\mathcal{H}}

\newcommand{\Wc}{\mathcal{W}}
\ifodd 0
\newcommand{\yj}[1]{{\color{blue}#1}}
\else
\newcommand{\yj}[1]{#1}
\fi

\ifodd 1
\newcommand{\yjc}[1]{{\color{red}(YJ: #1)}}
\else
\newcommand{\yjc}[1]{}
\fi

\begin{document}

\title{
When to Preempt? Age of Information Minimization under Link Capacity Constraint}
\author{Boyu~Wang, Songtao~Feng, and~Jing~Yang
\thanks{This work is presented in part in the 2018 IEEE International Workshop on Signal Processing Advances in Wireless Communications (SPAWC)~\cite{Boyu:SPAWC2018}. The authors are with the School of Electrical Engineering and Computer Science, The Pennsylvania State University, University Park, PA, 16802, USA. Email: \{bxw91, sxf302, yangjing\}@psu.edu.}}
 \maketitle

\begin{abstract}
In this paper, we consider a scenario where a source continuously monitors an object and sends time-stamped status updates to a destination through a rate-limited link. We assume updates arrive randomly at the source according to a Bernoulli process. Due to the link capacity constraint, it takes multiple time slots for the source to complete the transmission of an update. Therefore, when a new update arrives at the source during the transmission of another update, the source needs to decide whether to skip the new arrival or to switch to it, in order to minimize the expected average age of information (AoI) at the destination. We start with the setting where all updates are of the same size, and prove that within a broadly defined class of online policies, the optimal policy should be a renewal policy, and has a sequential switching property. We then show that the optimal decision of the source in any time slot has threshold structures, and only depends on the age of the update being transmitted and the AoI at the destination. We then consider the setting where updates are of different sizes, and show that the optimal Markovian policy also has a multiple-threshold structure. For each of the settings, we explicitly identify the thresholds by formulating the problem as a Markov Decision Process (MDP), and solve it through value iteration. Special structural properties of the corresponding optimal policy are utilized to reduce the computational complexity of the value iteration algorithm.
\end{abstract}


\section{Introduction}
\label{sec:introd}
Enabled by the proliferation of ubiquitous sensing devices and the pervasive wireless data connectivity, real-time status monitoring has become a reality in large-scale cyber-physical systems, such as power grids, manufacturing facilities, and smart transportation systems. However, the unprecedented high-dimensionality and generation rate of the sensing data also impose critical challenges on its timely delivery. In order to measure and ensure the freshness of the information available to the monitor, a metric called Age of Information (AoI) has been introduced and analyzed in various status updating systems~\cite{infocom/KaulYG12}. Specifically, at time $t$, the AoI in the system is defined as $t-y(t)$, where $y(t)$ is the time stamp of the latest received update at the destination. Since AoI depends on data generation as well as queueing and transmission, it is fundamentally different from traditional network performance metrics, such as throughput and delay.

Modeling the status updating process as a queueing process, time-average AoI has been analyzed in systems with a single server~\cite{infocom/KaulYG12,ciss/KaulYG12,isit/YatesK12,YatesK16,Pappas:2015:ICC,isit/NajmN16,isit/KamKNWE16,isit/ChenH16,Yates:Priority}, and multiple servers \cite{isit/KamKE13, isit/KamKE14,tit/KamKNE16,Yates:ParallelServers,Yates:Multicast}. Peak Age of Information (PAoI) has been introduced and studied in \cite{isit/CostaCE14,tit/CostaCE16,isit/HuangM15}. 
AoI minimization has also been investigated, either by controlling the generation process of the updates~\cite{infocom/SunUYKS16,SunPU17,isit/Yates15,ita/BacinogluCU15,Yang:2017:AoI,BacinogluU17,Ahmed:2018:ICC,Ahmed:2018:ITA,Uysal:2018:EH,Feng:2018:INFOCOM,Feng:2018:ISIT}, or by scheduling the transmission of updates that have already been generated~\cite{Modiano:2018:BC,Modiano:2018:MobiHoc,Hsu:2017:ISIT,Hsu:2018:ISIT,Kadota:2018:INFOCOM,He:TIT:2018}. Optimal status updating policy with knowledge of the server state has been studied in~\cite{infocom/SunUYKS16}. AoI-optimal sampling of a Wiener process is investigated in \cite{SunPU17}. Under an energy harvesting setting, optimal status updating have been studied in~\cite{isit/Yates15,ita/BacinogluCU15,Yang:2017:AoI,BacinogluU17,Ahmed:2018:ICC,Ahmed:2018:ITA,Uysal:2018:EH,Feng:2018:INFOCOM,Feng:2018:ISIT}. Transmission scheduling for AoI minimization has been studied for broadcast channels \cite{Modiano:2018:BC,Modiano:2018:MobiHoc,Hsu:2017:ISIT,Hsu:2018:ISIT}, and for multiple access channels \cite{Kadota:2018:INFOCOM}.
Age-optimal link scheduling in a multiple-source system with conflicting links is studied in~\cite{He:TIT:2018}, and the problem is shown to be NP-complete in general. 

Recently, a few works start to investigate the impact of service preemption on the time-average age in various status updating systems~\cite{ciss/KaulYG12,gamma,Bedewy:TIT:2019,Bedewy:ISIT:2017,Sun:2018:WKSP,mg,whentopreempt,Improve}. The common assumption is that new updates arrive at the source when it is in service (e.g., transmitting) of another update. Whether the source should preempt the current transmission thus would affect the age substantially. In \cite{ciss/KaulYG12}, it shows that for an $M/M/1$ system with a buffer, the last-come-first-served (LCFS) with preemption discipline achieves lower time-average age than LCFS without preemption. In \cite{gamma}, it indicates that when the service time follows a Gamma distribution, last-generated-first-served (LGFS) with preemption may not outperform LGFS without preemption. If the service times are i.i.d. and satisfy a New-Better-than-Used (NBU) distributional property, the non-preemptive LGFS policy is shown to be within a constant gap from the optimum age performance in \cite{Bedewy:TIT:2019}. Assuming exponential transmission time over network links, \cite{Bedewy:ISIT:2017} shows that a preemptive LGFS policy results in smaller age than any other causal policy. Similar optimal properties of the preemptive and non-preemptive LGFS policies have been extended to a multi-flow setting in \cite{Sun:2018:WKSP}.

When the transmitter does not have a buffer, the problem becomes whether to drop a new update arrival or to drop the unfinished update and starts the new one. In \cite{mg}, it studies an $M/G/1/1$ queue, and shows that when the updates are sent over an erasure channel, hybrid ARQ without preemption achieves smaller average AoI than the policy that always drops the unfinished updates.  In \cite{whentopreempt}, it considers an energy harvesting status updating system, and shows that if the service time is exponential and both the updating generation and energy harvesting process are Poisson, dropping unfinished updates achieves smaller age than dropping new arrivals when the system is in the ``energy rich" regime.  In \cite{Improve}, it focuses on stationary Markov and randomized policies that depend on the instantaneous AoI, and shows that whether to drop the new or old update depends on the service time distributions. 




In this paper, we investigate the age-optimal online transmission scheduling for a single link under the assumption that the link capacity is limited and each update takes multiple time slots to transmit. During the transmission of an update, new updates may arrive. We assume the size of each update is available to the source once it arrives, thus it has accurate information about how many time slots it takes to deliver the update. Then, without a buffer, the source has to decide whether to switch to the new arrival, or to continue its current transmission and skip the new update, based on causally known update arrival profile (size and arrival time). We consider two possible scenarios, based on different assumptions on the sizes of the updates:

1) {\it Updates of uniform size}. In this case, the transmission time of each update is fixed and the AoI will be reset to the same value once the update is delivered successfully. We first prove that within a broadly defined class of online policies, the optimal policy should be a renewal policy, and the decision-making over each renewal interval only depends on the arrival time of the updates in that interval. Then, we show that the optimal renewal policy has a multiple-threshold structure, which enables us to formulate the problem as an MDP, and identify the thresholds numerically through structured value iteration.

2) {\it Updates of non-uniform sizes}. In this case, each update may take a different number of time slots to transmit. Thus the AoI will be reset to different values when an update is delivered successfully. Therefore, the optimal policy depends on the arrival times of the updates, as well as their sizes. To make the problem tractable, we restrict to stationary Markov policies where the decision of the transmitter depends on the AoI at the destination, the age of the update being transmitted, and the size of the new update. We show that the optimal policy exhibits certain threshold structures along different dimensions of the system state. We then propose a structured value iteration algorithm to solve for the thresholds numerically.

The remainder of the paper is structured as follows: In Section \ref{sec:model}, we describe the system model and problem formulation. In Sections~\ref{sec:optimal} and~\ref{sec:nonuniform}, we consider the cases of uniform updates sizes and non-uniform update sizes, respectively. We evaluate the proposed policies numerically in Section~\ref{sec:simulation} and conclude in Section~\ref{sec:conclusion}.
\section{System Model and Problem Formulation} \label{sec:model}
We consider a single-link status monitoring system where the source keeps sending time-stamped status updates to a destination through a rate-limited link. We assume the time axis is discretized into time slots, labeled as $t=1, 2, 3,\cdots$. At the beginning of time slot $t$, an update packet is generated and arrives at the source according to an independent and identically distributed (i.i.d.) Bernoulli process $\{a_t\}$ with parameter $p$. We consider the scenario where the size of each update is large compared with the link capacity, so that it takes multiple time slots to transmit. The distribution of the transmission time of the updates will be specified later. Similar to \cite{Modiano:2018:BC,Hsu:2017:ISIT,Hsu:2018:ISIT}, we assume that at most one update can be transmitted during each time slot, and there is {\it no buffer} at the source to store any updates other than the one being transmitted. Therefore, once an update arrives at the source, it needs to decide whether to transmit it and drop the one being transmitted if there is any, or to drop the new arrival.

A status update policy is denoted as $\pi$, which consists of a sequence of transmission decisions $\{w_t\}$. We let $w_t\in\{0,1\}$. Specifically, when $a_t=1$, $w_t$ can take both values 1 and 0: If $w_t=1$, the source will start transmitting the new arrival in time slot $t$ and drop the unfinished update if there is one. We term this as {\it switch}; Otherwise, if $w_t=0$, the source will drop the new arrival, and continue transmitting the unfinished update if there is one, or be idle otherwise. We term this as {\it skip}. When $a_t=0$, we can show that dropping the update being transmitted is sub-optimal. Thus, we restrict to the policies under which $w_t$ can only take value 0, i.e., to continue transmitting the unfinished update if there is one, or to idle.

\begin{figure}[t]
	\centering
	\epsfxsize=6.5cm \epsfbox{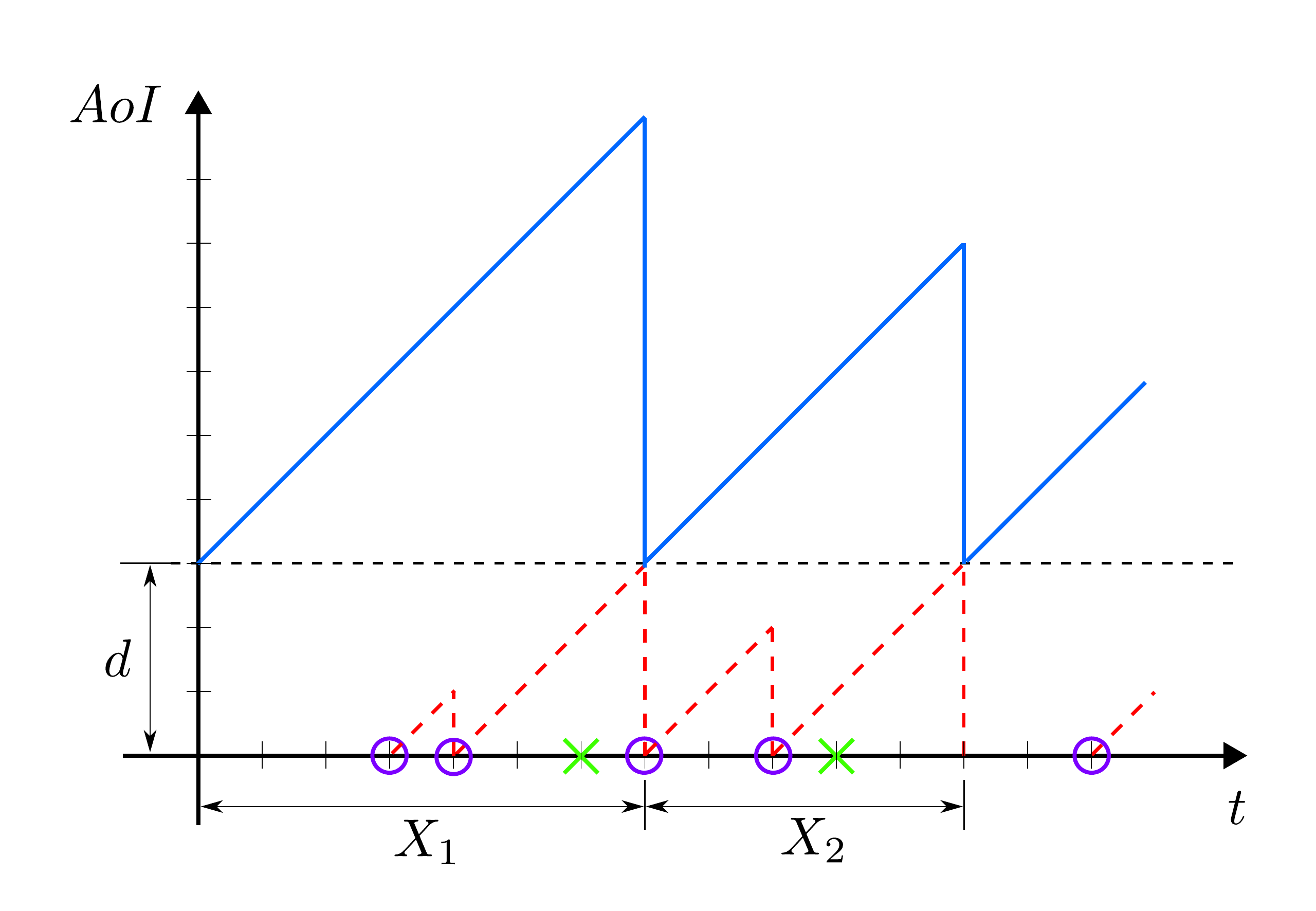}
	\vspace{-0.1in}
	\caption{\small{AoI evolution with equal transmission time $d_n=3$. Circles represent transmitted updates, and crosses represent skipped ones. Red dashed curve indicates the transmitted portion of the corresponding update. $S_1=7, S_2=12$.}} 
	\label{fig:AoI}
	\vspace{-0.1in}
\end{figure}

Let $S_n$ be the the time slot when an update is completely transmitted to the destination. Then,
the inter-update delays can be denoted as $X_n:=S_n-S_{n-1}$, for $n=1,2,\ldots$. Without loss of generality, we assume $S_0=0$. We note that since the update arrivals will either be dropped or transmitted immediately, the AoI after a completed transmission is reset to the transmission time of the delivered update, denoted as $d_n$. An example sample path of the AoI evolution under a given status update policy is shown in Fig.~\ref{fig:AoI}. 
As illustrated, some updates are skipped when they arrive, while others are transmitted partially or completely.

We use $N(T)$ to denote the total number of successfully delivered status updates over $(0,T]$. 
Define $R(T)$ as the cumulative AoI experienced by the system over $[0,T]$. Denote $R_n:=(2d_n+X_n)X_n/2$, i.e., the total AoI experienced by the receiver over the $n$th epoch $X_n$. Then,
\begin{align*}
R(T)&=\sum_{n=1}^{N(T)}R_n+ \frac{1}{2}(d_n+T-S_{N(T)})(T-S_{N(T)}).
\end{align*}
We focus on a set of {\it online} policies $\Pi$, in which the information available for determining $w_t$ includes the decision history $\{w_i\}_{i=1}^{t-1}$, the update arrival profile $\{a_i\}_{i=1}^{t}$, as well as the statistics of the update arrivals (i.e., arrival rate $p$ and the distribution of update sizes).
The optimization problem can be formulated as
\begin{eqnarray}\label{eqn:opt}
\underset{\pi\in\Pi}{\min}&\underset{T\rightarrow \infty}{\limsup} \quad \mathbb{E}\left[\frac{R(T)}{T}\right],\end{eqnarray}
where the expectation in the objective function is taken over all possible update arrival sample paths.

\section{Updates of Uniform Size}{\label{sec:optimal}}
In this section, we focus on the scenario where the updates are of the same size, and the required transmission time is equal to $d$ time slots, where $d$ is an integer greater than or equal to two.
\subsection{Structure of the Optimal Policy}
Consider the $n$th epoch, i.e., the duration between time slots $S_{n-1}+1$ and $S_n$ under any online policy in $\Pi$. Let $a_{n,k}$ be the time slot when the $k$th update after $S_{n-1}$ arrives, and let $x_{n,k}:=a_{n,k}-S_{n-1}$. Denote the update arrival profile in epoch $n$ as $\xv_n:=(x_{n,1},x_{n,2},\ldots)$. Then, we introduce the following definition.

\begin{Definition}[Uniformly Bounded Policy]
	Under an online policy $\pi\in\Pi$, if there exists a function $g(\xv)$, such that for any $\xv_n=\xv$, the length of the corresponding epoch $X_n$ is upper bounded by $g(\xv)$, $\Eb[g^2(\xv_n)]<\infty$, then this policy is a uniformly bounded policy.
\end{Definition}

Denote the subset of uniformly bounded policies as $\Pi'$. Then, we have the following theorem.
\begin{Theorem}\label{thm:renewal}
	The time-average AoI under any uniformly bounded policy $\pi\in\Pi'$ can always be improved by a renewal policy, under which $\{S_n\}_{n=1}^{\infty}$ form a renewal process, and the decision $\{w_t\}$ over the $n$th renewal epoch only depends on $\xv_n$ causally.
\end{Theorem}
The proof of Theorem~\ref{thm:renewal} is provided in Appendix~\ref{appx:thm:renewal}. Based on Theorem~\ref{thm:renewal}, in the following, we will focus on renewal policies that depend on $\xv_n$ only. 

\begin{Lemma}\label{lemma:idle}
	If the source is idle when an update arrives, it should start transmitting the update immediately.
\end{Lemma}
Lemma~\ref{lemma:idle} can be proved  through contradiction. If the source skips an update when it is idle, we can show that by switching to the new update instead, the AoI can always be improved statistically, thus the orignal policy cannot be optimal. The detailed proof is omitted for the brevity of this paper.
\if{0}
\begin{Proof}
	We proof this lemma through contradiction. We assume the under the optimal policy $\pi_0$, there exists a sample path under which in the $k^{th}$ epoch, the source skip the file arrives at time slot $i$ when the source is idle. For such sample paths, we will construct an alternative policy by let the source transmit the update at time slot $i$ and then follow policy $\pi_0$ after that. We note that if no updates arrive over time slots $[i+1,i+d-1]$, the source will finish the transmission at the end of time slot $i+d-1$ and reset the AoI to $d$. Compared with $\pi_0$, the AoI will be strictly improved under such sample paths by achieving a smaller AoI on average. Thus $\pi_0$ cannot be optimal.
\end{Proof}
\fi

\begin{Definition}[Sequential Switching Policy]
	A sequential switching (SS) policy is a renewal policy such that the source is allowed to switch to an update arriving at time slot $t$ \textbf{only if} it has switched to all updates that arrive before $t$ in the same epoch. 
\end{Definition}

\textbf{Remark:} The definition of SS policy implies that once a source skips a new update arrival at $t$, it will skip all of the upcoming update arrivals until it finishes the one being transmitted at $t$. We point out that an SS policy is in general different from threshold type of policies, as it does not impose any threshold structure on when the source should skip or switch to a new update arrival.
We have the following observations.

\begin{Lemma}\label{lemma:ssp}
	The optimal renewal policy in $\Pi'$ is an SS policy. 
\end{Lemma}
\begin{Lemma}\label{lemma:threshold}
	Consider a renewal epoch under the optimal SS policy in $\Pi'$. If the source switches to an update at the $i$th time slot in that epoch, then, there exists a threshold $\tau_i$, $\tau_i\geq i$, which depends on $i$ only, such that if the next update arrives before or at the $\tau_i$th time slot in that epoch, the source will switch to the new arrival; otherwise, it will skip any new updates until the end of the epoch.
\end{Lemma}

The proofs of Lemma~\ref{lemma:ssp} and Lemma~\ref{lemma:threshold} are provided in Appendix~\ref{appx:lemma:ssp} and Appendix~\ref{appx:lemma:threshold}, respectively. There are proved through contradiction, i.e., if the optimal policy does not exhibit the SS and threshold structures, respectively, we can always construct another renewal policy to achieve smaller average AoI. The alternative policy is constructed in a sophisticated manner to ensure that the first moment of the length of each renewal interval stays the same as under the original policy while strictly reducing the second moment, thus achieving a smaller expected average AoI.

Based on Lemma~\ref{lemma:ssp} and Lemma~\ref{lemma:threshold}, the structure of the optimal renewal policy is characterized in the following theorem.
\begin{Theorem}\label{thm:threshold}
	Consider a renewal epoch under the optimal SS policy in $\Pi'$. There exists a sequence of thresholds $\tau_1\geq\tau_2\geq \cdots \geq \tau_K$, such that if the source switches to an update at the $i$th time slot in that epoch, $1\leq i \leq  K$, and the next update arrives before or at the $\tau_i$th time slot in the epoch, the source will switch to the new arrival; Otherwise, if the next update arrives after the $\tau_i$th time slot, or the update being transmitted arrives after the $K$th slot in the epoch, the source will skip all upcoming arrivals until the end of the epoch.
\end{Theorem}
The proof of Theorem~\ref{thm:threshold} can be found in Appendix~\ref{appx:thm:threshold}.
It indicates that the optimal decision of the source in the $n$th epoch only depends on two parameters: the arrival time of the update being transmitted, and the arrival time of the new update, both relative to $S_{n-1}$, the end of the previous renewal epoch. 

\subsection{MDP formulation}{\label{subsec:mdp}}
Motivated by the Markovian structure of the optimal policy in Theorem~\ref{thm:threshold}, we cast the problem as an MDP, and numerically search for the optimal thresholds $\tau_1,\tau_2, \cdots \tau_K$ and $K$ as follows.

\textbf{States:} We define the state $\sv_t:=(\delta_t, u_t, a_t)$, where $\delta_t$ and $u_t$ are the AoI in the system, and the age of the unfinished update, at the beginning of time slot $t$, respectively. Thus, for state $\sv_t$, the correspond optimal threshold is $\tau_{\delta_t-d-u_t+1}$. $a_t\in\{0,1\}$ is the update arrival status. Then, $\delta_t\geq d$, $0\leq u_t\leq d-1$, and the state space $\Sc$ can be determined accordingly.

\textbf{Actions:} $w_t\in\{0,1\}$, as defined in Sec.~\ref{sec:model}. 

\textbf{Transition probabilities:} Denote transition probability from state $\sv_t$ to another state $\sv_{t+1}$ under action $w_t$ as $\pi(\sv_{t+1}|\sv_t,w_t)$. Then, if $w_t=0$, i.e., the transmitter either continues its transmission, or stays idle, we have
\begin{align}\label{eqn:keep}
\delta_{t+1}&=\left\{\begin{array}{ll}
\delta_t+1, & \mbox{if }u_t<d-1 \\
d, & \mbox{if }u_t=d-1
\end{array}
\right.,\\
{u_{t+1}}&=\left\{\begin{array}{ll}
u_t+1, & \mbox{if }  0<u_t<d-1 \\
0, & \mbox{otherwise} 
\end{array}
\right..\label{eqn:l}
\end{align}
If $w_t=1$, i.e., the transmitter switches to a new update arriving at the beginning of time slot $t$, then $\delta_{t+1}=\delta_t+1$, $u_{t+1}=1$.

\textbf{Cost:} Let $C(\sv_t, w_t)$ be the instantaneous age under state $\sv_t$, i.e., $C(\sv_t, w_t)=\delta_t$. 

Denote the relative value function as $V(\sv)$. Then the optimal Markovian policy to minimize the long-term average AoI is the solution to the following Bellman's equation~\cite{bertsekas2005dynamic}
\begin{align}\label{eqn:belman}
J+ V(\sv)&=\min_{w\in\Wc (\sv)} C(\sv;w)+ \Eb[V(\sv')|\sv,w].
\end{align}
where $J$ is the optimal average age, $\sv'$ is the next system state when $w$ is taken under state $\sv$, and $\Wc(\sv)$ is the set of allowable actions for given state $\sv$. $\Wc(\sv)=\{0,1\}$ if there is a new update arrival, and $\Wc(\sv)=\{0\}$ otherwise.


Let the reference state be $\sv_0:=(d,0,0)$. 
Then, the optimal policy can be determined through relative value iteration as follows:
\begin{align}\label{eqn:value iteration} 
V_{n+1}(\sv)\hspace{-0.04in}=\hspace{-0.1in}\min_{w\in\Wc(\sv)} \hspace{-0.03in} C(\sv, w)\hspace{-0.03in}+\hspace{-0.03in}\sum_{s'} \pi(\sv'|\sv,w)V_n(\sv')\hspace{-0.03in}-\hspace{-0.03in}V_n(\sv_0),
\end{align}
where $V_0(\sv)=0$, $\forall \sv$, and $V_n(\sv)$ converges to $V(\sv)$ as $n\rightarrow \infty$~\cite{bertsekas2005dynamic}. 

To make the problem numerically tractable, we truncate the state space of the original MDP as $\Sc_m:=\{\sv\in \Sc: \delta\leq \delta_{max}\}$ by capping the AoI $\delta$ at $\delta_{max}$, i.e., $[\delta]^+=\min{(\delta,\delta_{max})}$. It can be shown that when $\delta_{max}$ is sufficiently large, the optimal policy for the truncated MDP is identical to that of the original MDP \cite{Sennott1997}.

In order to further reduce the computational complexity, we leverage the multi-threshold structure of the optimal policy during the value iteration procedure, as detailed in the structured value iteration algorithm in Algorithm \ref{algorithm:ssa}.
With the multiple-threshold structure, Algorithm \ref{algorithm:ssa} does not need to seek the optimal action by equation (\ref{eqn:value iteration}) for all states in each iteration as the traditional value iteration algorithm does. Specifically, if the optimal action for a state $(\delta',u,1)$ is to {\it skip} the new arrival, the optimal action for state $(\delta, u ,1)$, $\delta>\delta'$, must be to skip as well. Similarly, if the optimal action for a state $(\delta,u',1)$ is to switch to the new arrival, the optimal action for state $(\delta, u ,1)$, $u<u'$, must be to switch. Thus, the computational complexity can be significantly reduced. Besides, since the truncated MDP is a unichain, Algorithm~\ref{algorithm:ssa} converges in a finite number of iterations~\cite{puterman2014markov}.

\begin{algorithm}[t]
	\caption{Structured Value Iteration: Uniform Update Size}\label{algorithm:ssa}
	\begin{algorithmic}[1]
		\State Initialize: $V_0(\sv)=0,\forall \sv\in \Sc_m$.
		\For{$i=0 : n$}
		\For{$\forall \sv\in \Sc_m$}
		\If{$a=0$}
		\State $w^*(\sv)=0$;
		\ElsIf{$\exists \delta'<\delta, w^*(\delta', u, 1)=0$}
		\State{$w^*(\sv)=0$};
		\ElsIf{$\exists u'<u, w^*(\delta, u', 1)=1$}
		\State{$w^*(\sv)=1$};
		\Else
		\State $\small{w^*(\sv)\hspace{-0.03in}=\hspace{-0.03in}\underset{w\in\{0, 1\}}{\arg\min}\hspace{0.02in}C(\sv,w)\hspace{-0.02in}+\hspace{-0.04in}\underset{\sv'}{\sum}\hspace{0.01in} \pi(\sv'|\sv,w)\hspace{0.01in}V_i(\sv')}$
		\EndIf
		\State $\small{V_{i+1}(\sv)\hspace{-0.03in}=\hspace{-0.03in}C(\sv; w^*(\sv))\hspace{-0.03in}+\hspace{-0.03in}\underset{\sv'}{\sum} \pi(\sv'|\sv,w)V_i(s')\hspace{-0.04in}-\hspace{-0.05in}V_i(\sv_0)}$
		\EndFor 
		\EndFor
		\State \textbf{return} $\{w^*(\sv)\}, \{V(\sv)\}.$
	\end{algorithmic}
\end{algorithm}
\section{Updates of Non-uniform Sizes}\label{sec:nonuniform}
In this section, we consider the scenario where the size of each update is non-uniform, and the required transmission time is a random variable following a known distribution. Specifically, we assume the required transmission time of each update is an i.i.d random variable following a probability mass function (PMF) $f_b$ over a bounded support in $\Zb_+$. For ease of exposition, we assume each update takes at least two time slots to transmit. Once a new update arrives at the source, its size is revealed to the transmitter immediately. Our objective is to decide whether to switch to the new update or to skip it, based on causal observations and the statistics of the update arrivals.

Compared with the uniform update size case, the source should also track the various sizes of the updates and take these into consideration when making decisions. In order to make the problem tractable, we restrict to Markovian polices where the decision at any time slot $t$ depends on the AoI at the destination, denoted as $\delta_t$, the remaining time to the destination of the update being transmitted, denoted as $l_t$, the required transmission time of the update being transmitted, denoted as $c_t$, and the required transmission time of the new arrival, denoted as $b_t$. If no new update arrives at the time slot $t$, $b_t=0$; Otherwise $b_t$ follows distribution $f_b$.

Denote the system state at time slot $t$ as $\sv_t:=(\delta_t, l_t, c_t, b_t)$. We note that $0\leq l_t\leq c_t-1$. Then, after the source takes an action $w_t\in\{0,1\}$, the state at time slot $t+1$ will change as follows. If $w_t=0$, the source either continues its transmission and skips the new update, or stays idle. Thus,
\begin{align}\label{eqn:keep}
\delta_{t+1}&=\left\{\begin{array}{ll}
\delta_t+1, & \mbox{if }l_t\neq 1, \\
c_t, & \mbox{if }l_t=1,
\end{array}
\right.\\
l_{t+1}&=\left\{\begin{array}{ll}
l_t-1, & \mbox{if }l_t>1, \\
0, & \mbox{if }l_t=0,1,
\end{array}
\right.\\
c_{t+1}&=\left\{\begin{array}{ll}
c_t, & \mbox{if }l_t>1, \\
0, & \mbox{if } l_t=0,1.
\end{array}
\right.
\end{align}

If $b_t\neq 0$ and $w_t=1$, i.e., the source switches to the new arrival, $ \delta_{t+1}=\delta_t+1$, $l_{t+1}=b_t-1$, $c_{t+1}=b_t$.

Let $C(\sv_t; w_t)$ be the immediate cost under state $\sv_t$ with action $w_t$. Similar to the uniform size case, we let $C(\sv_t; w_t)=\delta_t$.

\subsection{Structural Properties of Optimal Policy}
In order to obtain some structural properties of the optimal policy, in this section, we will first introduce an infinite horizon $\alpha$-discounted MDP as follows:
\begin{align}\label{eqn:alphabelman}
V^{\alpha}(\sv)&=\min_{w\in\Wc(\sv)} C(\sv;w)+ \alpha\Eb[V^{\alpha}(\sv')|\sv,w],
\end{align}
where $0<\alpha<1$. It has been shown that the optimal policy to minimize the average AoI can be obtained by solving (\ref{eqn:alphabelman}) when $\alpha\rightarrow 1$ \cite{bertsekas2005dynamic}. In order to identify the structural properties of the optimal policy, we start with the following value iteration formulation: 
\begin{align}\label{eqn:vi}
V^{\alpha}_{n+1}(\sv)&=\min_{w\in\Wc(\sv)} C(\sv;w)+ \alpha\Eb[V^{\alpha}_n(\sv')|\sv,w],
\end{align}
where $V^{\alpha}_0(\sv)=0, \forall\sv$.

Assume $\sv:=(\delta,l,c,b)$. If $b\neq 0$, denote 
\begin{align}
Q_n(\sv;w)&:=C(\sv;w)+\alpha\Eb[V^{\alpha}_n(\sv')|\sv,w],\\
\Delta Q_n^{w;w'}(\sv)&:=Q_n(\sv;w)-Q_n(\sv;w').
\end{align}
Then, $V^{\alpha}_{k+1}(\sv)=\min_{w\in\{0,1\}} Q_k(\sv;w)$. We have
\begin{align}\label{eqn:w0}
Q_k(\sv; 0)&=\left\{\begin{array}{ll}
\delta+\alpha\Eb[V^{\alpha}_k(\delta+1, {(l-1)^+}, c, \bar{b})], & \mbox{if }l\neq 1, \\
\delta+\alpha\Eb[V^{\alpha}_k(c, 0, 0,\bar{b})], & \mbox{if }l=1,
\end{array}\right.
\end{align}
and 
\begin{align}
Q_k(\sv; 1)&=\delta+\alpha\Eb[V^{\alpha}_k(\delta+1, b-1,b, \bar{b})].\label{eqn:w1}
\end{align}
where the expectation is taken with respect to $\bar{b}$, and $(x)^+=\max\{x,0\}$. We adopt $(l-1)^+$ in (\ref{eqn:w0}) to indicate that if $l=c=0$, and $w=0$, the system will remain idle. 

If $b=0$, $V^{\alpha}_{k+1}(\sv)=Q_k(\sv;0)$, where $Q_k(\sv;0)$ follows the same form of (\ref{eqn:w0}).


We have the following observations:
\begin{Lemma}\label{Lemma:nond}
Let $\sv:=(\delta,l,c,b)$. Then, $V^{\alpha}_n(\sv)$ is monotonically increasing in $\delta$, in $c$ for {$c>0$}, and in $b$ for $b>0$, $\forall n$.
\end{Lemma}
\begin{Proof}
	We first prove the monotonicity in $\delta$ through induction. First, it holds when $n=0$. Assume it holds for $n=k$. Then, it suffices to show that it holds for $n=k+1$ as well. If $b>0$, $V^{\alpha}_{k+1}(\sv)=\min_{w\in\{0,1\}} Q_k(\sv;w)$. We note that $Q_k(\sv;1)$ is an increasing function in $\delta$ according to the induction assumption. If $l=1$, $Q_k(\sv;0)$ is increasing in $\delta$ based on its definition in (\ref{eqn:w0}). If $l\neq1$, $Q_k(\sv;0)$ is increasing in $\delta$ according to the induction assumption. Then, $V^{\alpha}_{k+1}(\sv)=\min_{w} Q_k(\sv;w)$ must monotonically increase in $\delta$, as taking the minimum preserves the monotonicity. If $b=0$, $V^{\alpha}_{k+1}(\sv)=Q_k(\sv;0)$ is monotonically increasing in $\delta$ as well.

Next, we prove the monotonicity in $c$ for $c>0$. We prove it through induction as well. It holds when $n=0$. Assume it is true for $n= k$. We note that $Q_k(\sv; 1)$ is independent of $c$ when $b>0$. Besides, if $l=1$, $Q_k(\sv; 0)$ is increasing in $c$ based on the monotonicity of $V^{\alpha}_n(\sv)$ in $\delta$. If $1<l<c$, $Q_k(\sv; 0)$ is increasing in $c$ based on to the induction assumption. Thus $V^{\alpha}_{k+1}(\sv)$ is increasing in $c$ for $c>0$.

In order to show the monotonicity of $V^{\alpha}_n(\sv)$ in $b$ for $b>0$, we first show that $Q_n(\sv; 1)$ is increasing in $b$ for $b>0$, $\forall n$.	
Based on the definition of $Q_n(\sv; 1)$ in (\ref{eqn:w1}), it suffices to show $V^{\alpha}_n(\delta, b-q, b, \bar{b})$ is increasing in $b$, for $ 1\leq q\leq b$.  We note it holds when $n=0$. Assume $V^{\alpha}_k(\delta,b-q,b,\bar{b})$ is increasing in $b$. Then, we will show that $V^{\alpha}_{k+1}(\delta,b-q,b,\bar{b})$ is increasing in $b$ as well. We note that if $\bar{b}\neq 0$, $Q_k((\delta,b-q,b,\bar{b});1)$ is independent with $b$. Besides, when $q=b-1$, $Q_k((\delta,b-q,b,\bar{b});0)=\delta+\alpha\Eb[V^{\alpha}_k(b, 0, 0, \bar{b})]$, which is increasing in $b$ according to the monotonicity of $V^{\alpha}_k(b, 0, 0, \bar{b})$; when $ 1\leq q<b-1$, $Q_k((\delta,b-q,b,\bar{b});0)=\delta+\alpha\Eb[V^{\alpha}_k(\delta+1, b-q-1, b, \bar{b})]$, which is increasing in $b$ based on the induction assumption. Thus, $V^{\alpha}_{k+1}(\delta,b-q,b,\bar{b})$ is increasing in $b$ after taking the minimum of $Q_k((\delta,b-q,b,\bar{b});0)$ and $Q_k((\delta,b-q,b,\bar{b});1)$. The monotonicity of $Q_{n}(\sv; 1)$ in $b$ for $b>0$ is thus established.

Since $Q_{n}(\sv; 0)$ is independent of $b$, while $Q_{n}(\sv; 1)$ is increasing in $b$ when $b>0$, after taking the minimum of them, $V^{\alpha}_n(\sv)$ is increasing in $b$ for $b>0$ as well.
\end{Proof}

\begin{Lemma}\label{Lemma:induction1}
	$\Eb[V^{\alpha}_n(\delta,l_1,c_1,\bar{b})]\leq\Eb[V^{\alpha}_n(\delta,l_2,c_2,\bar{b})]$ if and only if $Q_{n-1}((\delta,l_1,c_1,.);0)\leq Q_{n-1}((\delta,l_2,c_2,.);0)$, $\forall n\geq1$.
\end{Lemma}	
\begin{Proof}
	First, we note that
	\begin{align}\label{eqn:expectation} 
	&\Eb[V^{\alpha}_n(\delta,l,c,\bar{b})]\nonumber\\
	&=(1-p)Q_{n-1}((\delta,l,c,0);0)\nonumber\\
	&\quad+p \Eb\left[\min_{w\in\{0,1\}} Q_{n-1}((\delta,l,c,\bar{b});w)\middle |\bar{b}\neq 0\right].
	\end{align}
	Besides, for $\bar{b}\neq 0$,
	\begin{align} 
	Q_{n-1}((\delta,l_1,c_1,\bar{b});1)=Q_{n-1}((\delta,l_2,c_2,\bar{b});1).
	\end{align}
	If 
	\begin{align} 
	Q_{n-1}((\delta,l_1,c_1,.);0)\leq Q_{n-1}((\delta,l_2,c_2,.);0),
	\end{align}
	then, according to (\ref{eqn:expectation}), 
	\begin{align} 
	\Eb[V^{\alpha}_n(\delta,l_1,c_1,\bar{b})]\leq\Eb[V^{\alpha}_n(\delta,l_2,c_2,\bar{b})].
	\end{align}
	The sufficiency is thus established.
	
	On the other hand, if
	\begin{align} 
	Q_{n-1}((\delta,l_1,c_1,.);0)> Q_{n-1}((\delta,l_2,c_2,.);0),
	\end{align}
	then for every possible $\bar{b}\neq 0$, we must have 
	\begin{align} 
	\min_{w\in\{0,1\}}{Q_{n-1}((\delta,l_1,c_1,\bar{b});w)}\geq	\min_{w\in\{0,1\}}{Q_{n-1}((\delta,l_2,c_2,\bar{b});w)}.\nonumber
	\end{align}
	Based on (\ref{eqn:expectation}) and the assumption that $0<p<1$, we have
	\begin{align} 
	\Eb[V^{\alpha}_n(\delta,l_1,c_1,\bar{b})]>\Eb[V^{\alpha}_n(\delta,l_2,c_2,\bar{b})],
	\end{align}
	which proves the necessity of the condition.
\end{Proof}

\begin{Corollary}\label{Lemma:induction2}
	For states with $l_1,l_2>1$, $\Eb[V^{\alpha}_n(\delta,l_1,c_1,\bar{b})]\leq\Eb[V^{\alpha}_n(\delta,l_2,c_2,\bar{b})]$ if and only if $\Eb[V^{\alpha}_{n-1}(\delta+1,l_1-1,c_1,\bar{b})]\leq\Eb[V^{\alpha}_{n-1}(\delta+1,l_2-1,c_2,\bar{b})]$, $\forall n\geq1$.
\end{Corollary}
\begin{Proof}
	Since $Q_{n-1}((\delta,l,c,.);0)=\delta+\alpha\Eb[V^{\alpha}_{n-1}(\delta+1,l-1,c,\bar{b})]$ when $l>1$, 
	$Q_{n-1}((\delta,l_1,c_1,.);0)\leq Q_{n-1}((\delta,l_2,c_2,.);0)$ is equivalent to
	\begin{align*}
	\Eb[V^{\alpha}_{n-1}(\delta+1,l_1-1,c_1,\bar{b})]\leq \Eb[V^{\alpha}_{n-1}(\delta+1,l_2-1,c_2,\bar{b})].
	\end{align*}
	The corollary is thus proved based on Lemma~\ref{Lemma:induction1}.
\end{Proof}

\begin{Corollary}\label{corollary:induction}
	For states with $l_1>1$, $\Eb[V^{\alpha}_n(\delta,l_1,c_1,\bar{b})]\leq\Eb[V^{\alpha}_n(\delta,0,0,\bar{b})]$ if and only if $\Eb[V^{\alpha}_{n-1}(\delta+1,l_1-1,c_1,\bar{b})]\leq\Eb[V^{\alpha}_{n-1}(\delta+1,0,0,\bar{b})]$, $\forall n\geq1$.
\end{Corollary}
Corollary~\ref{corollary:induction} can be proved in a way similar to the proof of Corollary~\ref{Lemma:induction2}, and is thus omitted.

Based on Lemmas~\ref{Lemma:nond}, \ref{Lemma:induction1} and Corollaries~\ref{Lemma:induction2},~\ref{corollary:induction}, we have the following theorem.
\begin{Theorem}\label{thm:non_uniform}
Denote $\sv=(\delta,l,c,b)$ and assume $b>0$. The optimal policy for the $\alpha$-discounted MDP has the following structure:
\begin{itemize}
\item[(a)] If the optimal action for $\sv$ is to switch, then the optimal action for any state $(\delta,l,c,b')$, $0<b'<b$, is to switch as well.
\item[(b)] If $0<b\leq l$, then the optimal action for $\sv$ is to switch to the new update.
\item[(c)] If the optimal action for $\sv$ is to switch, then for any $c'>c$, the optimal action for state $\sv'\triangleq(\delta,l,c',b)$ is to switch as well.
 \item[(d)] If $l=c=0$, then the optimal action for for $\sv$ is to switch to the new update.
 \item[(e)] If the optimal action for $\sv$ is to skip, then the optimal action for state $(\delta+1,l,c,b)$ is to skip as well. 
\end{itemize}
\end{Theorem}	

The proof of Theorem~\ref{thm:non_uniform} is provided in Appendix~\ref{appx:thm:non_uniform}. Theorem~\ref{thm:non_uniform}(a)  indicates the threshold structure on the size of the new update arrival, i.e., the source prefers to switch to a new update if its size is small, and will skip it if its size is large. Theorem~\ref{thm:non_uniform}(b) is a consequence of Theorem~\ref{thm:non_uniform}(a). Theorem~\ref{thm:non_uniform}(c) shows that there exists a threshold on the size of the update being transmitted, i.e., the source prefers to drop updates with larger sizes and switch to new updates. Theorem~\ref{thm:non_uniform}(d) says that the source should immediately start transmitting the new update arrival if it has been idle, which is consistent with Lemma~\ref{lemma:idle} for the uniform update size case. Theorem~\ref{thm:non_uniform}(e) essentially indicates the threshold structure on the instantaneous AoI at the destination: that the source prefers to skip new updates when the AoI is large, as it is in more urgent need to complete the current transmission and reset the AoI to a smaller value.

We point out that all of the structural properties of the optimal policy derived for the $\alpha$-discounted problem hold when $\alpha\rightarrow 1$\cite{Hsu:2017:ISIT}. Thus, the optimal policy for the time-average problem also exhibits similar structures.

 \subsection{Structured Value Iteration}
Following an approach similar to the uniform update size case, we leverage the structural properties of the optimal policy and develop a structured value iteration algorithm to obtain the thresholds numerically. The detailed algorithm is presented in Algorithm \ref{algorithm:ssa}.

\if{0}
Similar with the uniform $d$ case, we define $\delta_m$ as the boundary AoI, and truncate the state space of the original MDP as $\Sc_m=\{\sv\in \Sc: \delta\leq \delta_m\}$ to reduce the computational complexity.

\begin{Theorem}\label{Theorem:truncated}
	Let $J^*$ and $J_m^*$ be the optimal average cost of the MDP and truncated MDP, respectively. Then, $J_m^*\rightarrow J^*$ as $m\rightarrow\infty$.
\end{Theorem}
\begin{Proof}
	According to \cite{Sennott1997}, in order to show $J_m^*\rightarrow J^*$, it suffices to show that the following two conditions are satisfied:
	
	(1) There exists a nonnegative $L$, a nonnegative finite function $F(.)$ on $\sv$, such that $-L\leq V^{\alpha(m)}(\sv)-V^{\alpha(m)}(0)\leq F(\sv)$ for all $\sv\in \Sc_m$, $m=1,2,\cdots$, $0<\alpha<1$. (2) $\limsup_{m\rightarrow\infty}J_m^*\leq J^*$.
	
	We first show the first condition is satisfied. Let $C_{\sv,0}(\pi)$ and $C_{\sv,0}^{(m)}(\pi)$ be the expected cost from state $\sv$ to the reference state $0$ by applying the algorithm $\pi$ to the MDP and the truncated MDP, respectively. We considering $\pi$ as the myopic policy that only transmit $d_{\max}$ without preemption, where $d_{\max}$ is the maximum transmitting time of the arrivals. Then we know the average age is finite using the similar idea in our special case of uniform $d$. Thus, we know $C_{\sv,0}(\pi)<\infty$ according to Proposition 4 and $-C^{(m)}_{0,\sv}(\pi)\leq V^{\alpha(m)}(\sv)-V^{\alpha(m)}(0)\leq C_{\sv,0}^{(m)}(\pi)$ according the proof of Proposition 5 in~\cite{Sennott1997}. Besides, we can prove $C_{\sv,0}^{(m)}(\pi)\leq C_{\sv,0}(\pi)$ ($-C_{0,\sv}^{(m)}(\pi)\geq -C_{0,\sv}(\pi)$) using the similar technical in \cite{Hsu:2017:ISIT}. Thus, we can choose $F(\sv)=C_{\sv,0}(\pi)$. On the other hand, we know $V^{\alpha(m)}(\sv)-V^{\alpha(m)}(0)\geq -C_{0,\sv}^{(m)}(\pi)\geq-C_{0,\sv}(\pi)$. Moreover, as $V^{\alpha}(s)$ is non-decreasing in $\delta$, $c$, and $b$ (Lemma~\ref{Lemma:nond}-\ref{Lemma:nond3}), we can choose the reference state as $(\delta_0,l_0,c,b)=\min_{0\leq l_0\leq b_{\max}-1, 0\leq c\leq b_{max}, 0\leq b\leq b_{\max}}V^{\alpha}(\delta_0,l_0,c,b)$, where $\delta_0\geq 2*b_{max}$. Thus for all $\delta>\delta_0$, we know $V^{\alpha}(s)>V^{\alpha}(0)$. Only for the states with $\delta<\delta_0$ can probably result in lower value of $V^{\alpha}(s)$ than $V^{\alpha}(0)$. Therefore, we can choose $L=\max_{\sv\in\Sv:\delta<\delta_0}C_{(0,s)}(\pi)$.
	
	For the second condition, we will prove for the $\alpha$ discounted cost, we have $V^{\alpha(m)}(\sv)\leq V^{\alpha}(\sv)$, then the condition follows as
	\begin{align}
	J_m^*\hspace{-0.02in}=\hspace{-0.02in}\limsup_{\alpha\rightarrow 1}(1-\alpha)V^{\alpha(m)}\hspace{-0.02in}\leq\hspace{-0.02in}\limsup_{\alpha\rightarrow 1}(1-\alpha)V^{\alpha}\hspace{-0.02in}=\hspace{-0.02in}J^*,
	\end{align}
	according to \cite{bertsekas2005dynamic}.
	
	We prove $V^{\alpha(m)}(s)\leq V^{\alpha}(s)$ through induction. Denote $\delta(\sv)$ be the age at state $\sv$, according to the definition of the truncated MDP, we know when $\delta(\sv)<m$, $\delta(\sv')\leq m$ or $\delta(\sv)=m, \delta(\sv')<m$ we have $\pi^{(m)}(\sv'|\sv)=\pi(\sv'|\sv)$. When $\delta(\sv)=\delta(\sv')=m$, we have $\sum_{\sv\in\Sv_m} \pi^{(m)}(\sv'|\sv)=\sum_{\sv\in\Sv,\delta(\sv'')>m,l(\sv'')=l(\sv'),c(\sv'')=c(\sv'),b(\sv'')=b(\sv')}\pi(\sv'|\sv)$.
	
	When $n=0$. we have $V^{\alpha(m)}_0(s)\leq V^{\alpha}_0(s)$. Assume $V^{\alpha(m)}_k(s)\leq V^{\alpha}_k(s)$, then we have
	\begin{align}
	V^{\alpha(m)}_{k+1}(\sv)&=\hspace{-0.025in}\min_{w\in\{0,1\}}C(\sv;w)\hspace{-0.025in}+\hspace{-0.025in}\alpha\sum_{\sv'\in\Sc_m}\pi^{(m)}(\sv'|\sv,w)V^{\alpha(m)}_k(\sv')\nonumber\\
	&\leq\min_{w\in\{0,1\}}C(\sv;w)\hspace{-0.025in}+\hspace{-0.025in}\alpha\sum_{\sv'\in\Sc_m}\pi^{(m)}(\sv'|\sv,w)V^{\alpha}_k(\sv')\nonumber\\
	&\leq\min_{w\in\{0,1\}}C(\sv;w)\hspace{-0.025in}+\hspace{-0.025in}\alpha\sum_{\sv'\in\Sc}\pi(\sv'|\sv,w)V^{\alpha}_k(\sv')\nonumber\\
	&=V^{\alpha}_{k+1}(\sv),
	\end{align}
	where the first inequality results from the induction assumption, the second inequality results from
	\begin{align}
	\sum_{\sv'\in\Sc_m}\pi^{(m)}(\sv'|\sv,w)V^{\alpha}(\sv')\leq\sum_{\sv'\in\Sc}\pi(\sv'|\sv,w)V^{\alpha}(\sv'),
	\end{align}
	as $V^{\alpha}(\sv)$ is non-decreasing in $\delta$ according to Lemma~\ref{Lemma:nond}, and the structure of $\pi^{(m)}(\sv'|\sv,w)$.
\end{Proof}

We adapt the traditional relative value iteration algorithm by applying the structural properties:
\begin{align}\label{eqn:RVIA}
V_{n+1}(\sv)=\min_{w\in\{0,1\}}C(\sv;w)+\Eb[V_n(\sv')]-V_n(0),
\end{align}
Instead of updating $V(\Sv)$ for all states by minimizing (\ref{eqn:RVIA}), we first find if the structural properties hold, then we can determine an optimal action immediately. Otherwise, we find an optimal action according to Line $13$. Then we can reduce the computational complexity by reducing the minimum operation.
\fi

\begin{algorithm}[h]
	\caption{Structured Value Iteration for Non-uniform Update Size Case.}\label{algorithm:ssd}
	\begin{algorithmic}
		\State Initialize: $V_0(\sv)=0,\forall \sv\in \Sc_m$.
		\For{$i=0 : n$}
		\For{$\forall \sv\in \Sc_m$}
		\If{$d=0$}
		\State $w^*(\sv)=0$;
		\ElsIf{$l, c=0$}
		\State{$w^*(\sv)=1$};
		\ElsIf{$\exists \delta'<\delta, w^*(\delta', l, c, b)=0$}
		\State{$w^*(\sv)=0$};
		\ElsIf{$\exists c'\yj{>}c, w^*(\delta, l, c', b)=0$}
		\State{$w^*(\sv)=0$};
		\ElsIf{$\exists d'\yj{>}b, w^*(\delta, l, c, b')=1$}
		\State{$w^*(\sv)=1$};
		\Else
		\State $w^*(\sv)\hspace{-0.01in}=\hspace{-0.01in}\underset{w\in\{0, 1\}}{\arg\min}\hspace{0.01in}C(\sv;w)\hspace{-0.01in}+\hspace{-0.01in}\underset{\sv'}{\sum}\hspace{-0.01in} \pi(\sv'|\sv,w)V_i(\sv')$
		\EndIf
		\State $V_{i+1}(\sv)\hspace{-0.04in}=\hspace{-0.04in}C(\sv, w^*(\sv))\hspace{-0.02in}+\hspace{-0.02in}\underset{\sv'}{\sum} \pi(\sv'|\sv,w)V_i(\sv')\hspace{-0.03in}-\hspace{-0.03in}V_i(\sv_0)$
		\EndFor 
		\EndFor
		\State \textbf{return} $\{w^*(\sv)\}, \{V(\sv)\}.$
	\end{algorithmic}
\end{algorithm}
\section{Numerical Results}{\label{sec:simulation}}
In this section, we numerically search for the optimal policies for both uniform and non-uniform update size cases according to Algorithm~\ref{algorithm:ssa} and Algorithm~\ref{algorithm:ssd}, respectively. For both truncated MDPs, we set $\delta_{max}=1000$ and the number of iterations to be $10,000$. 

\subsection{Updates of Uniform Size}
\begin{figure}[t]
	\centering
	\begin{minipage}[t]{4.3cm}
		\centering
		\centerline{\includegraphics[width=4.2cm]{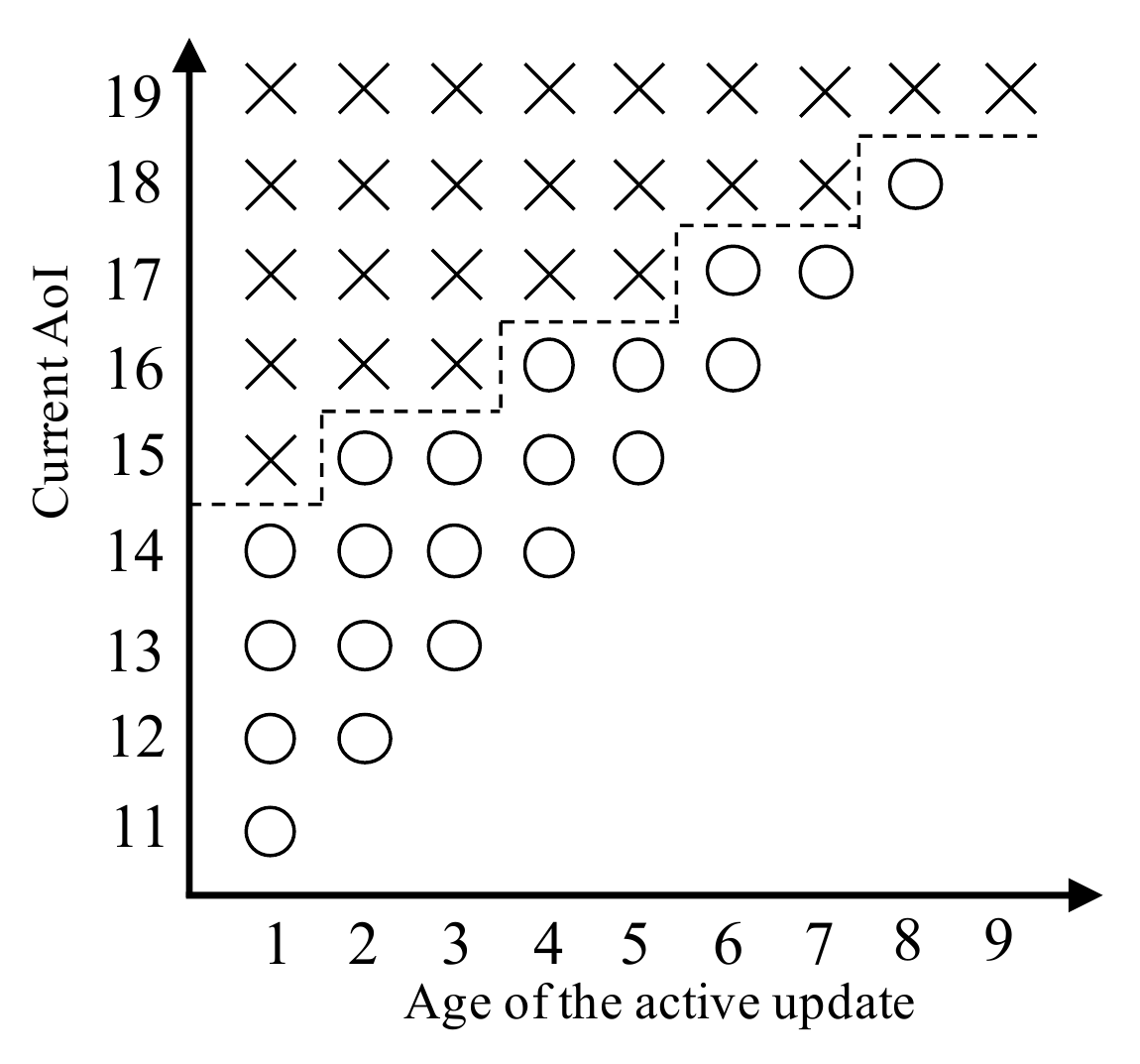}}
		\vspace{-0.05in}
		\centerline{\small{(a)}}
		\label{fig:mdp}	
	\end{minipage}
	\begin{minipage}[t]{4.2cm}
		\centering
		\centerline{\includegraphics[width=4.3cm]{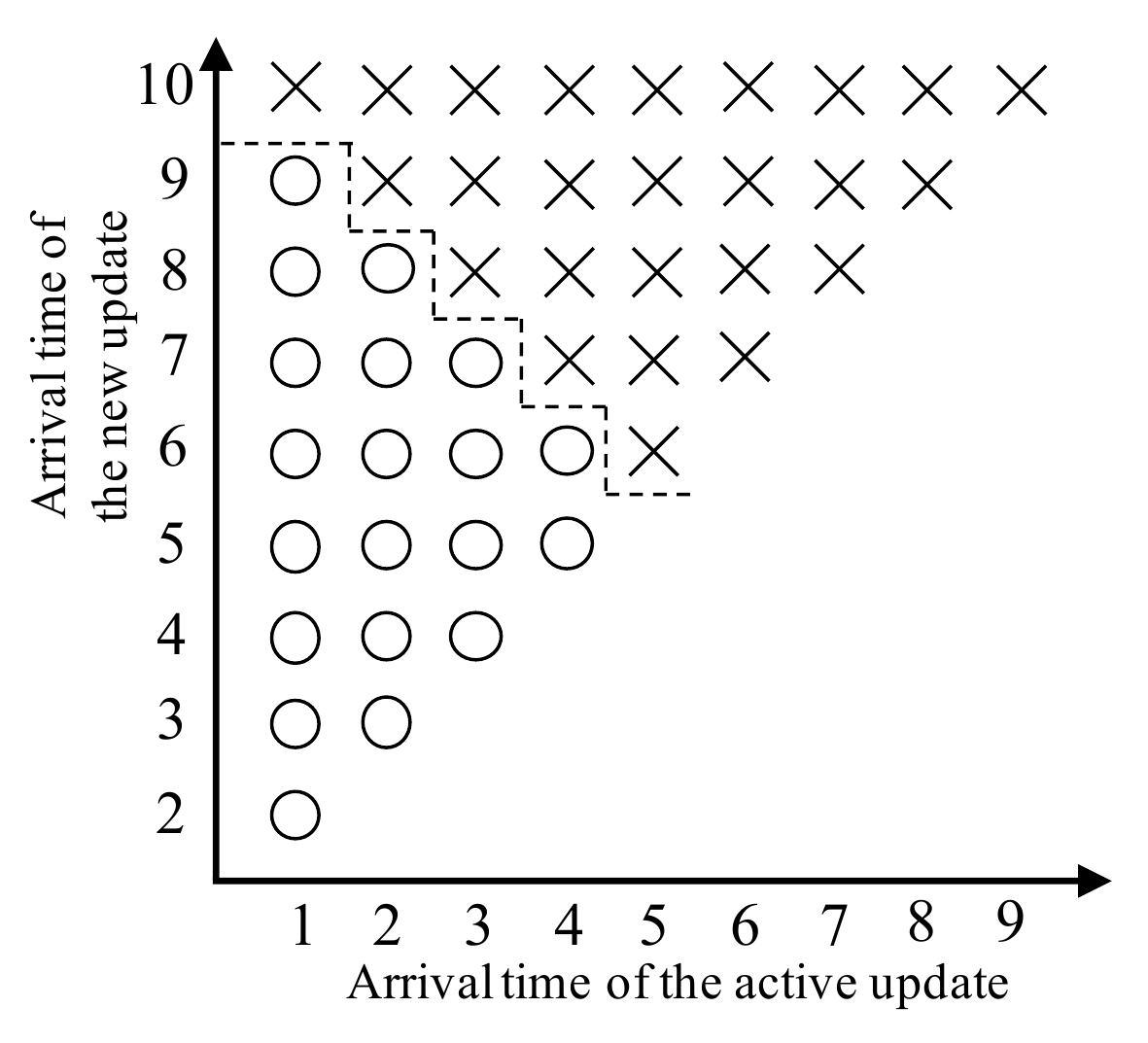}}
		\vspace{-0.05in}
		\centerline{\small{(b)}}
		\label{fig:amdp}
	\end{minipage}
	\caption{The optimal policy when $p=0.07$, $d=10$. Circles represent {\it switch}, while crosses represent {\it skip}.}
	\label{fig:switch}
	\vspace{-0.1in}
\end{figure}

First, we focus on the uniform update size case. We set $d=10$, $p=0.07$. Fig.~\ref{fig:switch}(a) shows the optimal action for each state $(\delta,u,1)$. We note the monotonicity of the thresholds in both $\delta$ and $u$. We then plot the optimal action for each pair of arrival time of the update being transmitted (i.e., active update) and that of the new arrival in a renewal epoch in Fig.~\ref{fig:switch}(b). We note that the thresholds $\tau_1=9$, $\tau_2=8$, $\tau_3=7$, and $\tau_4=6$. They are monotonically decreasing, as predicted by Theorem~\ref{thm:threshold}. When the update being transmitted arrives later than the third time slot in that epoch, all upcoming updates will be skipped. 

\begin{figure}[t]
	\vspace{-0.1in}
	\centering
   \epsfxsize=8cm \epsfbox{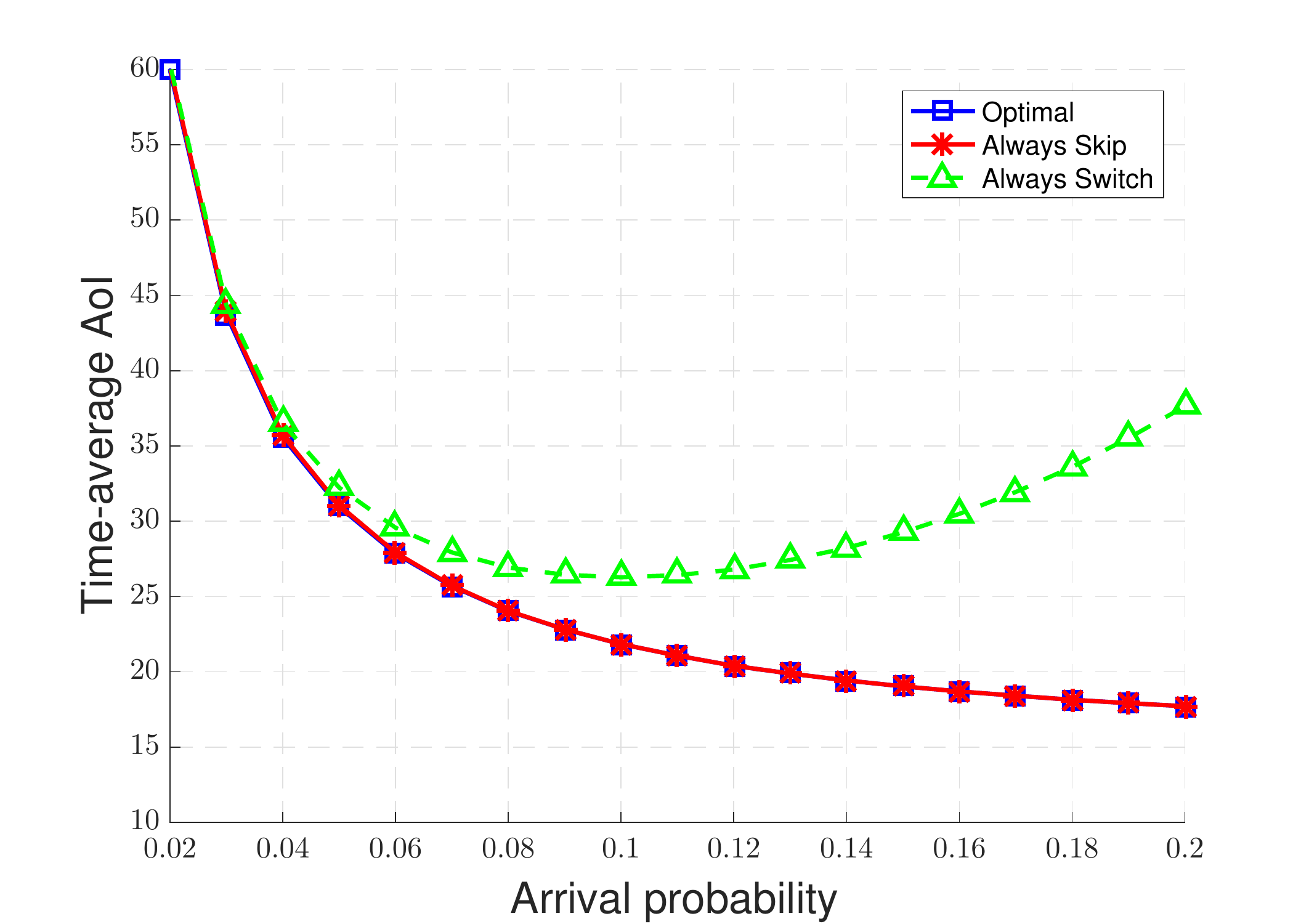}
   	\vspace{-0.05in}
	\caption{Average AoI with $d=10$.}
		\vspace{-0.1in}
	\label{fig:d10}
\end{figure}

\begin{figure}[t]
	\vspace{-0.1in}
	\centering
	\epsfxsize=8cm \epsfbox{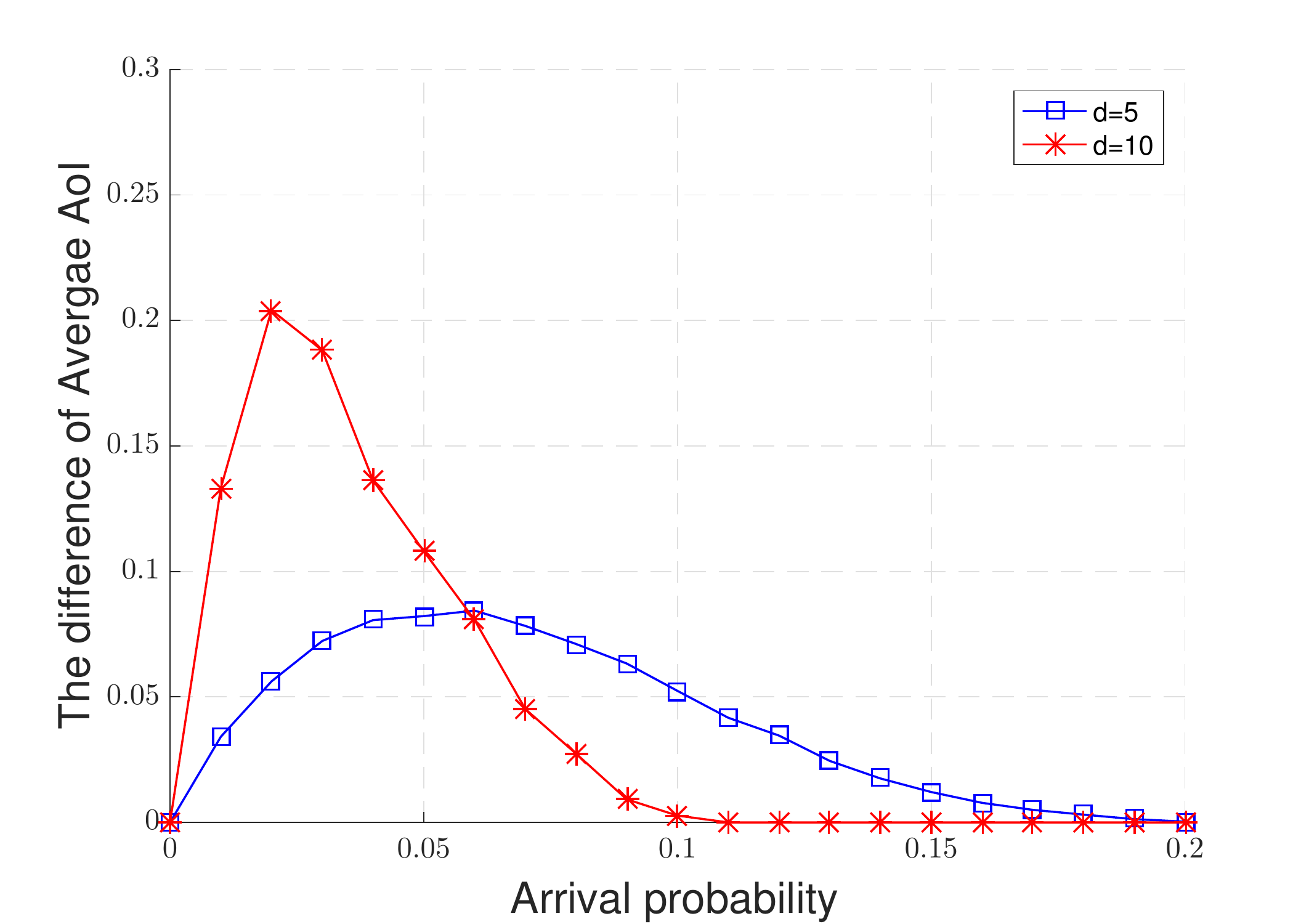}
	\vspace{-0.05in}
	\caption{Performance gap between the optimal policy and Always Skip.}
	\label{fig:ageResults2}
	\vspace{-0.1in}
\end{figure}

Then, we evaluate the time-average AoI under the optimal policy identified by Algorithm~\ref{algorithm:ssa} and two baseline policies, namely, Always Skip and Always Switch policies, over $10,000$ time slots. Under the Always Skip policy, the source will never switch to any new update arrival until it finishes the one being transmitted, while under the Always Switch policy, the source will always switch to new updates upon their arrivals. As we observe in  Fig.~\ref{fig:d10}, the performance differences between those three policies are negligible when $p$ is close to $0$. This is because when $p$ is small, the source won't receive a new update before it finishes transmitting the current update with high probability, thus the source behaves almost identically under all three policies.  As $p$ increases, the performances of the optimal policy and the Always Skip policy are still very close to each other, while the Always Switch policy renders the highest AoI. 
To distinguish the performances of the Always Skip policy and the optimal policy, we plot the the performance gap between them on time-average AoI in Fig.~\ref{fig:ageResults2}. As we note, as $p$ increases, the performance gap first increases and then decreases to zero. This is because when $p$ is not very large, the source still needs to preempt the transmission of an update occasionally in order to regulate the inter-update delays under the optimal policy; and when $p$ is sufficiently large, this becomes unnecessary, as it will have a new update arrival soon after a successful update. Thus, the optimal policy becomes the same as the Always Skip policy in this regime. It is interesting to note the surprisingly small performance gap between the optimal policy and the Always Skip policy in the simulation results. Bounding it theoretically is one of our future steps.

\if{0}
\begin{figure}[t]
	\centering
	\epsfxsize=8cm \epsfbox{./graph/d5.eps}
	\caption{Average age under different probabilities with $d=5$.}
	\label{fig:d5}
\end{figure}
\fi

\subsection{Updates of Non-uniform Sizes}
In this subsection, we focus on the non-uniform update size case. For illustration, we consider a scenario where the updates are of two possible sizes, and the corresponding transmission times are 5 and 8 time slots, respectively. We assume $f_d$ is a uniform distribution over $\{5,8\}$, and {the probability of arrival $p=0.14$}.

\begin{figure}[t]
	\centering
	\begin{minipage}[t]{3.65cm}
		\centering
		\centerline{\includegraphics[height=4.5cm]{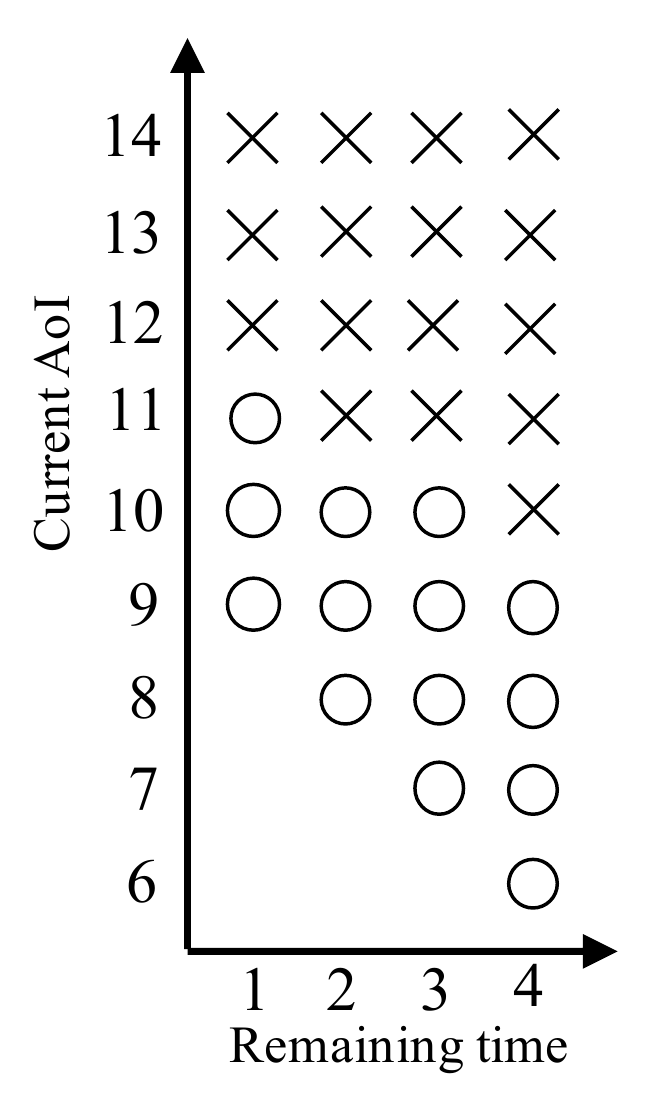}}
		\vspace{-0.05in}
		\centerline{\small{(a) $c=5$, $b=5$.}}
	\end{minipage}
	\begin{minipage}[t]{4.3cm}
		\centering
		\centerline{\includegraphics[height=4.5cm]{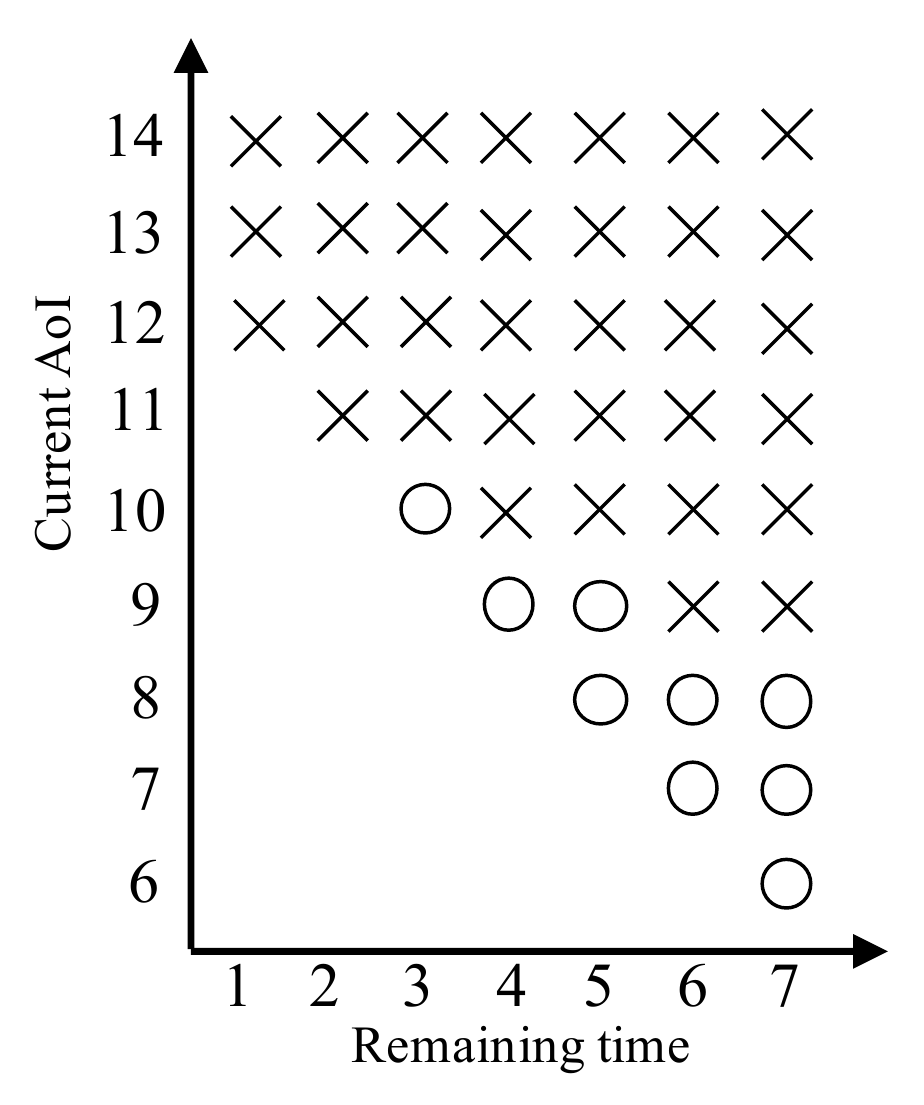}}
		\vspace{-0.05in}
		\centerline{\small{(b)} $c=8$, $b=8$.}
	\end{minipage}		
	\\	
	\begin{minipage}[t]{3.65cm}
		\centering
		\centerline{\includegraphics[height=4.5cm]{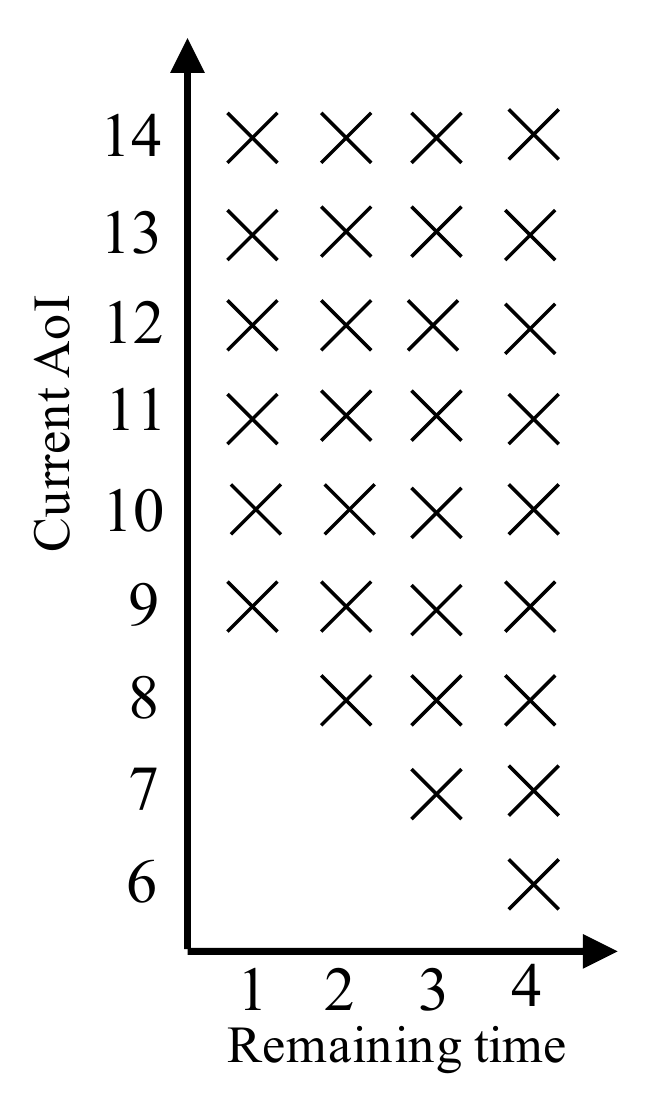}}
		\vspace{-0.05in}
		\centerline{\small{(c) $c=5$, $b=8$.}}
	\end{minipage}			
	\begin{minipage}[t]{4.3cm}
		\centering
		\centerline{\includegraphics[height=4.5cm]{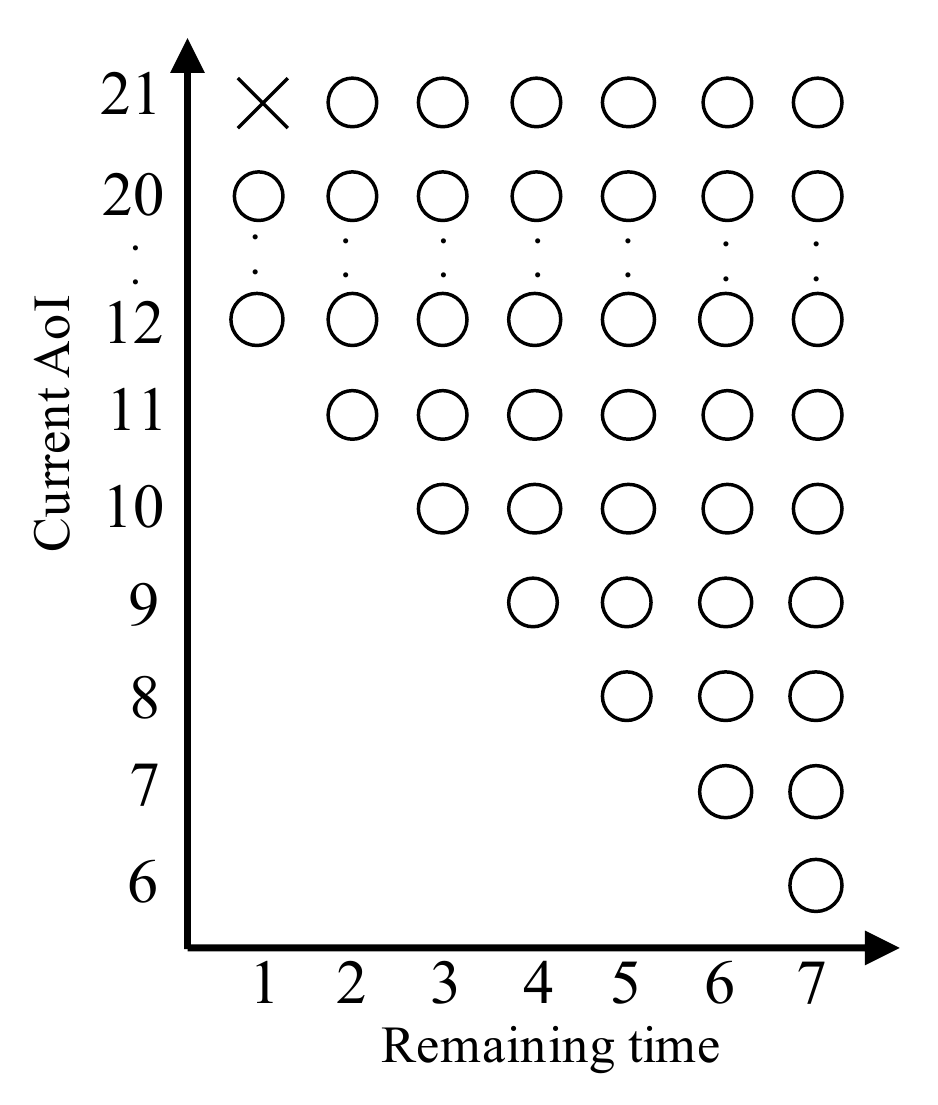}}
		\vspace{-0.05in}
		\centerline{\small{(d) $c=8$, $b=5$.}}
	\end{minipage}			
	\caption{The optimal policy when $p=0.14$. Circles represent {\it switch}, while crosses represent {\it skip}.}
	\label{fig:nonud}
	\vspace{-0.1in}
\end{figure}

We first obtain the the optimal policy according to Algorithm~\ref{algorithm:ssd}. The policy depends on the instantaneous state $\sv:=(\delta,l,c,b)$, i.e., the current AoI $\delta$, the remaining transmission time of the active update $l$, the sizes of the active update and the new update, $c$ and $b$, respectively. Thus, we plot the optimal policy for each fixed $(c,b)$ pair in Fig.~\ref{fig:nonud}. We note that the optimal policy exhibits the structural properties predicted in Theorem~\ref{thm:non_uniform}.
We also note that when $c=5,b=8$, the source always prefers to skip the new update. This can be explained by the intuition that switching to an update with longer transmission time will lead to larger AoI when $p$ is not very small.

Then, we evaluate the time-average AoI performances under the optimal policy identified by Algorithm~\ref{algorithm:ssd}, the Always Skip policy, and the Always Switch policy for different $p$. We note the optimal policy outperforms the other two policies. Compared with the uniform update size case, the performance gap between the optimal policy and the Always Skip policy is more significant. This indicates that switching to a new update of small size would make a substantial impact on the overall AoI. 


\begin{figure}[h]
	\centering
	\epsfxsize=8cm \epsfbox{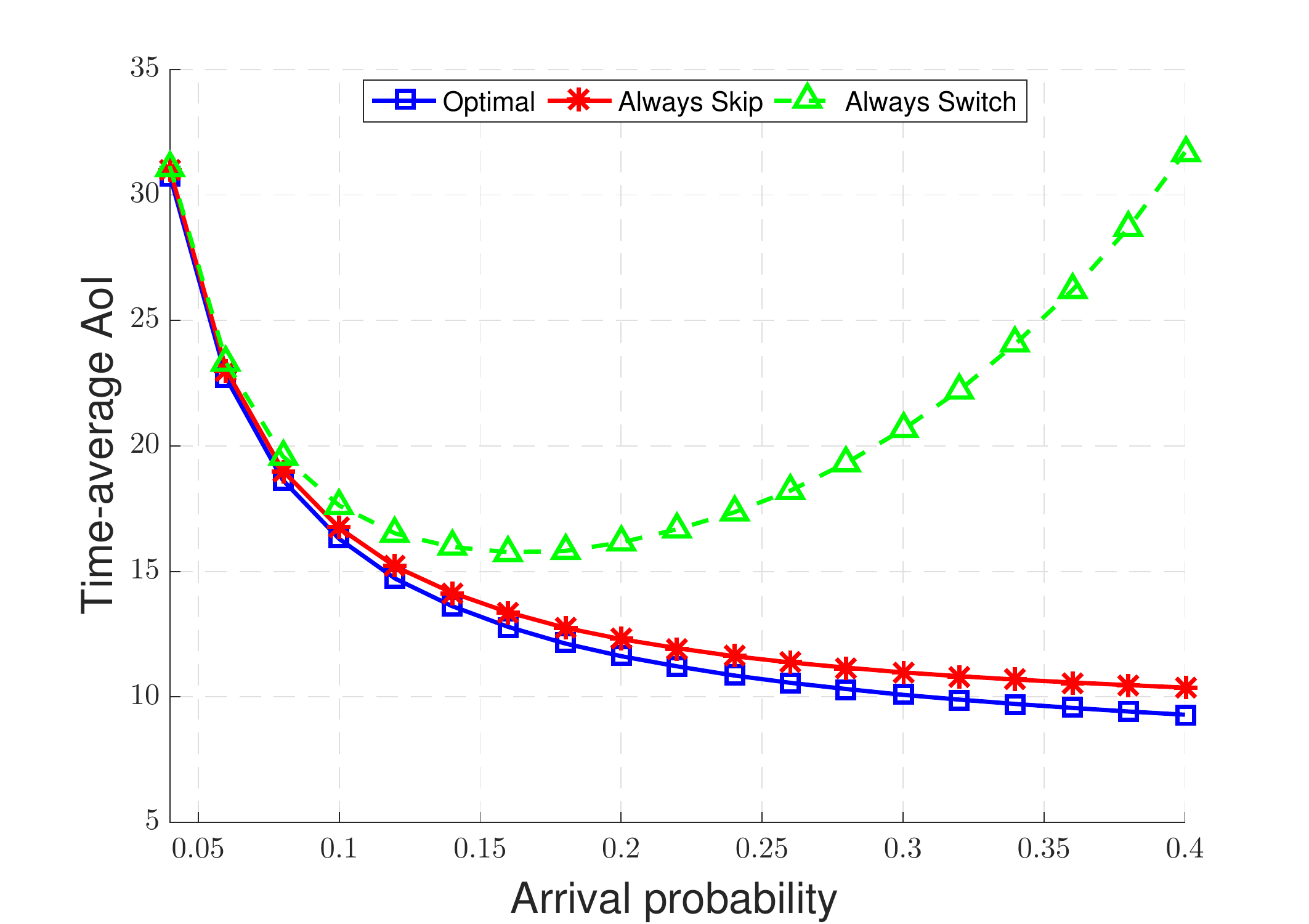}
	\caption{Average AoI comparison. $b\in\{5,8\}$.}
	\label{fig:d68}
\end{figure}
\allowdisplaybreaks
\section{Conclusions}{\label{sec:conclusion}}
In this paper we have considered a single-link status updating system under link capacity constraint. We first assumed the update size is uniform, and proved that within a broadly defined class of online policies, the optimal policy should be a renewal policy, and has a sequential switching property. We then showed that the optimal decision of the source in any time slot has a multiple-threshold structure, and only depends on the age of the update being transmitted and the AoI in the system. We then considered a more general case that the updates have different sizes, and showed the optimal policy exhibits certain threshold structure along different dimensions of the system state. For both cases, the optimal policies are numerically identified through structured value iteration under the MDP framework.

\appendices
\section{Proof of Theorem~\ref{thm:renewal}}\label{appx:thm:renewal}
\begin{Lemma}\label{lemma:bounded}
Under any $\pi\in\Pi'$, it must have $\lim_{T\rightarrow\infty}\frac{\Eb[X^2_{N(T)+1}]}{T}=0$.
\end{Lemma}

\begin{Proof}
	The proof of this lemma is adapted from the proof of Theorem~3 in \cite{Yang:2017:AoI}. For the completeness, we provided the detailed proof here.
	
	Denote $F_n(t)$ as the cumulative distribution function of $S_n$ under a uniform bounded policy, i.e., $F_n(t)=\Pb[S_n\leq t]$.
	Recall that $N(t)$ is the number of successfully delivered status updates over $(0,t]$. We have
	\begin{align}
	\Eb[N(t)]=\sum_{n=0}^{\infty} F_n(t)  \label{2eqn:extra-0}  .
	\end{align}
	We note that (\ref{2eqn:extra-1}) follows from the definition of uniformly bounded policy and (\ref{2eqn:extra-2}) follows from the fact that $g(\xv)$ is independent of other parameters. 
	\begin{align}
	&\Eb[X_{n+1}^2 \lv_{S_{n+1}>T} | S_n=t ]  \nonumber \\
	&=\Eb[X_{n+1}^2 \lv_{X_{n+1}>T-t} | S_n=t ]  \\
	&\leq \Eb_\xv [ g^2(\xv) \lv_{g(\xv)>T-t} | S_n=t, \xv_n=\xv  ] \label{2eqn:extra-1} \\
	&=\Eb_\xv [ g^2(\xv) \lv_{g(\xv)>T-t}  | \xv_n=\xv ] :=G(T-t)  \label{2eqn:extra-2} 
	\end{align}
	where (\ref{2eqn:extra-1}) follows from the definition of uniformly bounded policy and (\ref{2eqn:extra-2}) follows from the fact that $g(\xv)$ is independent of other parameters. We note that
	\begin{align}
	\lim_{\Delta \rightarrow \infty} G(\Delta) = 0  \label{2eqn:extra-13} .
	\end{align}
	We have
	\begin{align}
	&\Eb[X_{N(T)+1}^2]  \nonumber \\
	&=\sum_{n=0}^{\infty} \int_{0}^{T} \Eb[X_{n+1}^2 \lv_{S_{n+1}>T} | S_n=t ] dF_n(t)  \\
	&\leq \int_{0}^{T} G(T-t) d\left( \sum_{n=0}^{\infty}F_n(t) \right)  \label{2eqn:extra-3} \\
	&=\int_{0}^{T} G(T-t) d\Eb[N(t)]  \label{2eqn:extra-4}  ,
	\end{align}
	where (\ref{2eqn:extra-3}) follows from (\ref{2eqn:extra-2}), and (\ref{2eqn:extra-4}) follows from (\ref{2eqn:extra-0}).
	
	For any fixed $\Delta$ satisfying $0\leq \Delta \leq T$, we have
	\begin{align}
	&\frac{1}{T}\int_{0}^{T} G(T-t) d\Eb[N(t)] \nonumber \\
	&=\frac{1}{T} \int_{0}^{T-\Delta} G(T\hspace{-0.025in}-t) d\Eb[N(t)]  \hspace{-0.02in} + \hspace{-0.02in} \frac{1}{T}\hspace{-0.04in} \int_{T-\Delta}^{T} \hspace{-0.05in} G(T-t) d\Eb[N(t)] \nonumber \\
	&\leq G(\Delta)\frac{\Eb[N(T\hspace{-0.02in}-\hspace{-0.02in}\Delta)]}{T} \hspace{-0.02in}+\hspace{-0.02in} G(0)\frac{\Eb[N(T)]\hspace{-0.02in}-\hspace{-0.02in}\Eb[N(T\hspace{-0.025in}-\hspace{-0.025in}\Delta)]}{T}  \label{2eqn:extra-5} \\
	&\leq G(\Delta)\frac{T-\Delta}{dT}+G(0)\frac{\Delta}{dT}
	\label{2eqn:extra-6} ,
	\end{align}
	where (\ref{2eqn:extra-5}) follows from that fact that $G(t)$ is a non-increasing function, (\ref{2eqn:extra-6}) follows from the fact that the inter-update delay is greater than or equal to $d$ time slots.
	
Therefore, for any $\Delta\geq 0$
	\begin{align}
	\lim_{T\rightarrow \infty} \frac{1}{T}\int_{0}^{T} G(T-t) d\Eb[N(T)]  =\frac{G(\Delta)}{d}. \label{2eqn:extra-8}
	\end{align}
	Since $\lim_{\Delta \rightarrow \infty}G(\Delta)=0$ in (\ref{2eqn:extra-13}), we have $\lim_{T\rightarrow\infty}\frac{\Eb \left[X^2_{N(T)+1}\right]}{T}=0$.
\end{Proof}

Denote $\{X_n\}$ as the inter-update delays generated under a uniformly bounded policy $\pi\in\Pi'$. In general, the policy over the $n${th} epoch depends on the history over the previous $(n-1)${th} epoch, denoted as $\Hc^{n-1}$, the update arrivals over the $n${th} epoch, denoted as $\xv_n$, as well as external randomness, denoted as $\omega_n$. Let $R_n(\Hc^{n-1},\xv_n,\omega_n)$ be the cumulative AoI over the $n${th} epoch under the sample path for given $\Hc^{n-1}$, $\omega_n$, and $\xv_n$, and $X_n(\Hc^{n-1},\omega_n,\xv_n)$ be the corresponding length of the epoch. Then, we have the inequalities in (\ref{eqn:thm-1})-(\ref{eqn:thm-5}).
In (\ref{eqn:thm-5}), $R^*:=\min_{n,\Hc^{n-1},\omega_n}\frac{\Eb_{\xv_n}\left[R_n(\xv_n,\Hc^{n-1},\omega_n)\right]}{\Eb_{\xv_n} [X_n(\xv_n, \Hc^{n-1},\omega_n)]}$, i.e., the minimum average AoI over one epoch, among all possible epochs, history, as well as external randomness. We then apply this policy over all epochs irrespective history and the external randomness. Due to the memoryless property of the update arrival process, this is always feasible and achieves $R^*$ over each epoch. Thus, we can always obtain a renewal policy outperform the original $\pi$ which only causally depends on $\xv_n$.
\begin{table*}[h]
\begin{align}
&\lim_{T\rightarrow\infty}\frac{\Eb[R(T)]}{T}\geq\lim_{T\rightarrow\infty}\frac{\Eb\left[\sum_{n=1}^{N(T)}R_n\right]}{T}=\lim_{T\rightarrow\infty}\frac{\Eb\left[\sum_{n=1}^{N(T)+1}R_n\right]}{T} \label{eqn:thm-1}\\
&\geq \lim_{T\rightarrow\infty}\frac{\Eb\left[\sum_{n=1}^{N(T)+1}R_n\right]}{2\Eb[S_{N(T)+1}]}=\lim_{T\rightarrow\infty}\frac{ \sum_{n=1}^{\infty} \Eb\left[R_n\lv_{n\leq N(T)+1}\right]}{ \sum_{n=1}^{\infty}\Eb[X_n\lv_{n\leq N(T)+1}]}\\
&=\lim_{T\rightarrow\infty}\frac{ \sum_{n=1}^{\infty} \Eb_{\Hc^{n-1}}\left[\Eb_{\xv_n}\left[R_n(\xv_n,\Hc^{n-1},\omega_n)\right]\lv_{n\leq N(T)+1}|\Hc^{n-1},\omega_n \right]}{ \sum_{n=1}^{\infty}\Eb[X_n\lv_{n\leq N(T)+1}]}\\
&=\lim_{T\rightarrow\infty}\frac{ \sum_{n=1}^{\infty} \Eb_{\Hc^{n-1}}\left[\Eb_{\xv_n} [X_n(\xv_n, \Hc^{n-1},\omega_n)]\frac{\Eb_{\xv_n}\left[R_n(\xv_n,\Hc^{n-1},\omega_n)\right]\lv_{n\leq N(T)+1}}{\Eb_{\xv_n} [X_n(\xv_n, \Hc^{n-1},\omega_n)]} |\Hc^{n-1},\omega_n\right]}{\sum_{n=1}^{\infty}\Eb[X_n\lv_{n\leq N(T)+1}]}\\
&\geq\lim_{T\rightarrow\infty}\frac{ \sum_{n=1}^{\infty} \Eb_{\Hc^{n-1},\omega_n}\left[\Eb_{\xv_n} [X_n(\xv_n, \Hc^{n-1},\omega_n)] R^*\lv_{n\leq N(T)+1} |\Hc^{n-1},\omega_n\right]}{\sum_{n=1}^{\infty}\Eb[X_n\lv_{n\leq N(T)+1}]}= R^*\label{eqn:thm-5}
\end{align}
\end{table*}
\section{Proof of Lemma~\ref{lemma:ssp}}\label{appx:lemma:ssp}
We prove this lemma through contradiction. Now assume the optimal policy $\pi_0$ is not an SS policy. Without loss of generality, we consider the first renewal epoch starting at time 0 (the beginning of time slot $1$). We assume under $\pi_0$ there exists a sample path under which the source transmits the new update arrival at time slot $i$ and does not switch to the next arrival at time slot $j$ in the same epoch, i.e., $i<j<i+d$. Depending on the upcoming random arrivals, the sample path may evolve into different sample paths. Denote the set of such sample paths as $\Fc_j$, as they share the same history up to time slot $j$. We can partition $\Fc_{j}$ into two subsets:
\begin{itemize}
	\item $\Fc_{j,1}$: The source skips all the upcoming arrivals and completes the transmission of the update arrives at $i$.
	\item $\Fc_{j,2}$: The source switches to some later arrival.
\end{itemize}
Let $X^{\pi_0}$ be the corresponding length of the renewal epoch under policy $\pi_0$. Then, $X^{\pi_0}=i+d-1$ for sample paths in $\Fc_{j,1}$, and $X^{\pi_0}>j+d-1$ for sample paths in $\Fc_{j,2}$.

We now construct two policies $\pi_1$ and $\pi_2$ as follows. Under both $\pi_1$ and $\pi_2$, the source will behave exactly the same as under $\pi_0$ for all sample paths not in $\Fc_j$. However, for the sample paths in $\Fc_j$, the actions the source will take after $j$ will be different. Specifically, under $\pi_1$, the source will finish the update that arrives at time slot $i$ irrespective of other factors. Therefore, for all sample paths in $\Fc_j$ under $\pi_0$, the corresponding length of the renewal epoch under $\pi_1$ will be $X^{\pi_1}=i+d-1$ under $\pi_1$. For $\pi_2$, we will let the source first switch to the arrival at time slot $j$, and then switch to a later arrival whenever the source switches under $\pi_0$. Then, for the sample paths in $\Fc_{j,1}$ under $\pi_0$, the corresponding length of the renewal epoch will be changed to $X^{\pi_2}=j+d-1$ under $\pi_2$, while for those in $\Fc_{j,2}$,  $X^{\pi_2}=X^{\pi_0}$.

Therefore, considering all possible sample paths under those policies, we have
$\Eb[X^{\pi_1}]<\Eb[X^{\pi_0}]<\Eb[X^{\pi_2}]$,
which implies that there must exist a $\rho$, $0<\rho<1$, such that
\begin{align}\label{eqn:E[X]}
\rho \Eb[X^{\pi_1}]+(1-\rho)\Eb[X^{\pi_2}]=\Eb[X^{\pi_0}].
\end{align}

We will then construct a randomized policy $\pi'$, under which it follows $\pi_1$ with probability $\rho$ and follows $\pi_2$ with probability $1-\rho$. Apparently, the expected length of the renewal epoch under $\pi'$, denoted as $X^{\pi'}$, will be the same as that under $\pi_0$. 

Next, we will show that $\Eb[(X^{\pi'})^2]\leq \Eb[(X^{\pi_0})^2]$. Denote $\rho_1:=\frac{\Pb[\Fc_{j,1}]}{\Pb[\Fc_j]},\rho_{2}:=\frac{\Pb[\Fc_{j,2}]}{\Pb[\Fc_j]}$. Then, (\ref{eqn:E[X]}) can be expressed as
\begin{align}\label{eqn:E[X]_2}
&\rho(i+d-1)+(1-\rho)\left[(j+d-1)\rho_1+ \Eb[X^{\pi_0}|\Fc_{j,2}]\rho_{2}\right]\nonumber\\
&=(i+d-1)\rho_1+\Eb[X^{\pi_0}|\Fc_{j,2}]\rho_{2},
\end{align} 
which can be reduced to
\begin{align}\label{eqn:E[X]_3}
&(1-\rho)\rho_{1}(j+d-1)\nonumber\\
&=(\rho_1-\rho) (i+d-1)+\rho \rho_{2}\Eb[X^{\pi_0}|\Fc_{j,2}].
\end{align}
Since $\Eb[X^{\pi_0}|\Fc_{j,2}]>j+d-1$, $(1-\rho)\rho_{1}=(\rho_1-\rho)+\rho \rho_{2}$, (\ref{eqn:E[X]_3}) implies that
$\rho_1-\rho>0$.
Dividing both sides of (\ref{eqn:E[X]_3}) by $(1-\rho)\rho_{1}$, we have
\begin{align*}
j+d-1&=\frac{\rho_1-\rho}{(1-\rho)\rho_{1}}(i+d-1)+\frac{\rho \rho_{2}}{(1-\rho)\rho_{1}} \Eb[X^{\pi_0}|\Fc_{j,2}].
\end{align*}
Note that $\frac{\rho_1-\rho}{(1-\rho)\rho_{1}}$ and $\frac{\rho \rho_{2}}{(1-\rho)\rho_{1}}$ form a valid distribution. Therefore, based on Jensen's inequality, we have
\begin{align*}
& (j+d-1)^2\nonumber\\
& < \frac{\rho_1-\rho}{(1-\rho)\rho_{1}} (i+d-1)^2+\frac{\rho \rho_{2}}{(1-\rho)\rho_{1}} \left(\Eb[X^{\pi_0}|\Fc_{j,2}]\right)^2\nonumber\\
&\leq\frac{\rho_1-\rho}{(1-\rho)\rho_{1}} (i+d-1)^2+\frac{\rho \rho_{2}}{(1-\rho)\rho_{1}} \Eb\left[\left(X^{\pi_0}\right)^2|\Fc_{j,2}\right],
\end{align*}
which is equivalently to
\begin{align}\label{eqn:E[X^2]}
&\rho(i+d-1)^2+(1-\rho)\Big[(j+d-1)^2\rho_1+ \Eb[\left(X^{\pi_0}\right)^2|\Fc_{j,2}]\rho_{2}\Big]\nonumber\\
&<(i+d-1)^2\rho_1+\Eb[(X^{\pi_0})^2|\Fc_{j,2}]\rho_{2}.
\end{align} 
I.e.,
\begin{align}\label{eqn:E[X^2]_2}
\rho \Eb[(X^{\pi_1})^2]+(1-\rho)\Eb[(X^{\pi_2})^2]<\Eb[(X^{\pi_0})^2].
\end{align}
Combining (\ref{eqn:E[X]}) and (\ref{eqn:E[X^2]_2}), we have 
\begin{align}
\frac{1}{2}\frac{\rho \Eb[(X^{\pi_1})^2]+(1-\rho)\Eb[(X^{\pi_2})^2]}{\rho \Eb[X^{\pi_1}]+(1-\rho)\Eb[X^{\pi_2}]}< \frac{1}{2}\frac{\Eb[(X^{\pi_0})^2]}{\Eb[X^{\pi_0}]},
\end{align}
i.e., the new policy $\pi'$ achieves a lower expected average AoI than $\pi_0$, which contradicts with the assumption that $\pi_0$ is optimal.
\section{Proof of Lemma~\ref{lemma:threshold}}\label{appx:lemma:threshold}
We prove this lemma through contradiction. Again, we focus on the first renewal epoch that starts at time 0. Assume under the optimal SS policy $\pi_0$, the source transmits the update that arrives in time slot $i$. It will then skip the {\em next arrival} if it arrives at time slot $j$, and switches to it if it arrives at time slot $k$, with $j<k<i+d$. We aim to perturb $\pi_0$ a bit to obtain a new policy $\pi'$ and show that it achieves a lower expected average AoI. 

Similarly to the proof of Lemma~\ref{lemma:ssp}, we consider all sample paths under $\pi_0$ that choose to transmit the update that arrives at time slot $i$, while the next arrival time is either $j$ or $k$. Denote the subset of such sample paths as $\Fc_i$. Again, we divide $\Fc_i$ into two subsets.
\begin{itemize}
	\item $\Fc_{i,1}$: The next arrival time is $j$. Then, the sample paths will skip $j$ and finish transmitting the update arrives at $i$. The corresponding length of renewal interval $X^{\pi_0}=i+d-1$.
	\item $\Fc_{i,2}$: The next arrival time is $k$. The sample paths will switch to the new arrival, and the corresponding length of the renewal interval $X^{\pi_0}\geq k+d-1$.
\end{itemize}

Then, we construct the new policy $\pi'$ in this way: For all sample paths in $\Fc_{i,1}$, the new policy will let the source randomly switch to the next arrival at $j$ and finish transmitting it with probability $\rho_j$, and let the rest sample paths follow the original policy $\pi_0$, i.e., skip the new arrival and finish transmitting the update arrives at $i$. For all sample paths in $\Fc_{i,2}$, the new policy will let the source randomly skip the next arrival at $k$ with probability $\rho_k$, and let the rest sample paths follow the original policy $\pi_0$. Then, under $\pi'$, we have
\begin{align}
\Eb[X^{\pi'}|\Fc_{i,1}]&= \rho_j (j+d-1) +(1-\rho_j)(i+d-1),\\
\Eb[X^{\pi'}|\Fc_{i,2}]&= \rho_k (i+d-1) +(1-\rho_k)\Eb[X^{\pi_0}|\Fc_{i,2}].
\end{align}
Since $i+d+1<j+d+1<k+d+1$, there must exist a pair of $\rho_j$ and $\rho_k$ such that
\begin{align}
\Eb[X^{\pi'}|\Fc_i]&=\Eb[X^{\pi_0}|\Fc_i],  \label{eqn:lemma3-11}
\end{align}
i.e.,
\begin{align}
\rho_j (j-i) \Pb[\Fc_{i,1}]\hspace{-0.02in}=\hspace{-0.02in}\rho_k [\Eb[X^{\pi_0}|\Fc_{i,2}]\hspace{-0.03in}-\hspace{-0.02in} (i\hspace{-0.02in}+\hspace{-0.02in}d\hspace{-0.02in}-\hspace{-0.02in}1)]\Pb[\Fc_{i,2}].   \label{eqn:lemma3-1}
\end{align}
Dividing both side of Eqn.~(\ref{eqn:lemma3-1}) by $\rho_j\Pb[\Fc_{i,1}]$ and then adding $i+d-1$, we have
\begin{align}
j+d-1=&\frac{\rho_j \Pb[\Fc_{i,1}] \hspace{-0.03in} - \hspace{-0.03in} \rho_k \Pb[\Fc_{i,2}]}{\rho_j  \Pb[\Fc_{i,1}]}(i+d-1) \nonumber\\
&+ \frac{\rho_k \Pb[\Fc_{i,2}]}{\rho_j  \Pb[\Fc_{i,1}]} \Eb[X^{\pi_0}|\Fc_{i,2}]. \label{eqn:jensen}
\end{align}
Since $\Eb[X^{\pi_0}|\Fc_{i,2}]\geq k+d-1$ and $i<j<k$, based on Eqn.~(\ref{eqn:lemma3-1}), we have
\begin{align}
\frac{\rho_k\Pb[\Fc_{i,2}]}{\rho_j \Pb[\Fc_{i,1}]} = \frac{j-i}{\Eb[X^{\pi_0}|\Fc_{i,1}]-(i+d-1)} \leq \frac{j-i}{k-i} 
\leq 1.
\end{align}
Thus, $\frac{\rho_j \Pb[\Fc_{i,1}]- \rho_k\Pb[\Fc_{i,2}]}{\rho_j \Pb[\Fc_{i,1}]}$ and $\frac{\rho_k\Pb[\Fc_{i,2}]}{\rho_j \Pb[\Fc_{i,1}]}$ form a valid distribution.

Applying Jensen's inequality to Eqn.~(\ref{eqn:jensen}), we have
\begin{align}
(j+d-1)^2 \leq &\frac{\rho_j \Pb[\Fc_{i,1}]- \rho_k\Pb[\Fc_{i,2}]}{\rho_j \Pb[\Fc_{i,1}]}(i+d-1)^2\nonumber\\
&+ \frac{\rho_k\Pb[\Fc_{i,2}]}{\rho_j \Pb[\Fc_{i,1}]} \left( \Eb[X^{\pi_0}|\Fc_{i,2}]\right)^2 \\
\leq & \frac{\rho_j \Pb[\Fc_{i,1}]- \rho_k\Pb[\Fc_{i,2}]}{\rho_j \Pb[\Fc_{i,1}]}(i+d-1)^2\nonumber\\
&+ \frac{\rho_k\Pb[\Fc_{i,2}]}{\rho_j \Pb[\Fc_{i,1}]}  \Eb[(X^{\pi_0})^2|\Fc_{i,2}]. \label{eqn:lemma3-2}
\end{align}
Therefore,
\begin{align*}
&\Pb[\Fc_i]\Eb[(X^{\pi'})^2|\Fc_i]\nonumber\\
&=\Pb[\Fc_{i,1}]\left[ \rho_j(j+d-1)^2 + (1-\rho_j) (i+d-1)^2 \right]  \nonumber \\
& \quad+ \Pb[\Fc_{i,2}]\left[ \rho_k (i+d-1)^2 + (1-\rho_k) \Eb[(X^{\pi_0})^2|\Fc_{i,2}] \right]  \\
&\leq \Pb[\Fc_{i,1}] (i+d-1)^2 + \Pb[\Fc_{i,2}] \Eb[(X^{\pi_0})^2|\Fc_{i,2}] \\
&=\Eb[(X^{\pi_0})^2|\Fc_i]\Pb[\Fc_i]. 
\end{align*}

Thus policy $\pi'$ achieves lower expected average AoI, which contradicts with the assumption that $\pi_0$ is optimal. Therefore, if the source skips the next update if it arrives at time slot $j$, it must skip it if it arrives later than $j$. Picking the minimum $j$ ($j>i$) in which a skip would occur, denote it as $j^*$. Then, the threshold $\tau_i$ equals $j^*-1$. Since the proof does not depend other information except $i$, $\tau_i$ depends on $i$ only. 
\section{Proof of Theorem~\ref{thm:threshold}}\label{appx:thm:threshold}
Recall that we assume $d\geq 2$. If $d=2$, we have only one threshold $\tau_2$. Thus, in the following, we focus on the case $d>2$ with three possible scenarios, i.e., $\tau_i>i+1, \tau_i=i+1, \tau_i=i$.

We will start with the scenario that $\tau_i>i+1$. We aim to show the monotonicity of the switch thresholds, i.e., $\tau_{i+1}\leq \tau_i$. We prove it through contradiction. Assume under optimal policy $\pi_0$ there exists two subsets of sample paths as follows.

\begin{itemize}
	\item $\Fc_{i,\tau_i+1}$: The source switches to the update at time $i$ and the next arrival is at $\tau_i+1$. Then, under $\pi_0$, it will skip the arrival and finish transmitting the update arrives at $i$. Therefore, 
	\begin{align}
	\Eb[X^{\pi_0}|\Fc_{i,\tau_i+1}]=i+d-1. 
	\end{align}
	\item $\Fc_{i+1,\tau_i+1}$: The source switches to the update at time $i+1$ and the next arrival is at $\tau_i+1$. Since  $\tau_{i+1}> \tau_i$, under $\pi_0$, it will switch to the new arrival. We have 
	\begin{align}
	\Eb[X^{\pi_0}|\Fc_{i+1,\tau_i+1}]\geq \tau_{i}+d>i+d+1\label{eqn:T2_1}. 
	\end{align}
\end{itemize}
Then, we will modify $\pi_0$ to $\pi'$ as follows:
For sample paths in $\Fc_{i,\tau_i+1}$, the source will randomly switch to the new arrival at $\tau_i+1$ with probability $\rho_i$ and then follow policy $\pi_0$ afterwards. Note that the sample paths after switching at $\tau_i+1$ should be the same as those in $\Fc_{i+1,\tau_i+1}$ according to Lemma~\ref{lemma:threshold}. Therefore, we have
\begin{align}\label{eqn:T2_2}
\Eb[X^{\pi'}|\Fc_{i,\tau_i\hspace{-0.025in}+\hspace{-0.025in}1}]\hspace{-0.025in}=\hspace{-0.025in}\rho_i \Eb[X^{\pi_0}|\Fc_{i\hspace{-0.015in}+\hspace{-0.015in}1,\tau_i\hspace{-0.02in}+\hspace{-0.02in}1}]\hspace{-0.025in}+\hspace{-0.025in}(1\hspace{-0.025in}-\hspace{-0.025in}\rho_i) (i\hspace{-0.025in}+\hspace{-0.025in}d\hspace{-0.025in}-\hspace{-0.025in}1) .
\end{align}
For sample paths in $\Fc_{i+1,\tau_i+1}$, the source will randomly skip the new arrival at $\tau_i+1$ with probability $\rho_{i+1}$ and finish transmitting the arrival at $i+1$; It will stick with policy $\pi_0$ for the rest sample paths. Then
\begin{align}\label{eqn:T2_3}
&\Eb\hspace{-0.01in}[\hspace{-0.02in}X^{\pi'}\hspace{-0.02in}|\hspace{-0.02in}\Fc_{i\hspace{-0.01in}+\hspace{-0.01in}1,\tau_i\hspace{-0.01in}+\hspace{-0.01in}1}]\hspace{-0.04in}=\hspace{-0.04in}\rho_{i\hspace{-0.01in}+\hspace{-0.01in}1} (i\hspace{-0.04in}+\hspace{-0.04in}d)  
\hspace{-0.04in}+\hspace{-0.04in}(1\hspace{-0.04in}-\hspace{-0.04in}\rho_{i\hspace{-0.02in}+\hspace{-0.02in}1}) \Eb[X^{\pi_0}|\Fc_{i\hspace{-0.02in}+\hspace{-0.02in}1,\tau_i\hspace{-0.02in}+\hspace{-0.02in}1}]  .
\end{align} 
Denote $\rho_1:=\frac{\Pb[\Fc_{i,\tau_i+1}]}{\Pb[\Fc_{i,\tau_i+1}]+\Pb[\Fc_{i+1,\tau_i+1}]}$, $\rho_2:=1-\rho_1$. 
Let $\Fc=\Fc_{i,\tau_i+1}\cup\Fc_{i+1,\tau_i+1}$. We have
\begin{align}
\Eb[X^{\pi'}|\Fc]=&[\rho_i\rho_1+(1-\rho_{i+1})\rho_2]\Eb[X^{\pi_0}|\Fc_{i+1,\tau_i+1}]\nonumber\\
&+(1-\rho_i)(i+d-1)\rho_1+\rho_{i+1}(i+d)\rho_2.
\end{align} 
Based on Eqns.~(\ref{eqn:T2_1}), (\ref{eqn:T2_2}), and (\ref{eqn:T2_3}), we have 
\begin{align}
(i\hspace{-0.03in}+\hspace{-0.03in}d\hspace{-0.02in}-\hspace{-0.03in}1)\rho_1\hspace{-0.03in}+\hspace{-0.03in}(i\hspace{-0.02in}+\hspace{-0.02in}d)\rho_2\hspace{-0.02in}\leq \hspace{-0.02in}\Eb[X^{\pi'}|\Fc]\hspace{-0.02in}\leq\hspace{-0.02in}\Eb[X^{\pi_0}|\Fc_{i\hspace{-0.01in}+\hspace{-0.01in}1,\tau_i\hspace{-0.01in}+\hspace{-0.01in}1}].
\end{align}
By varying $\rho_i$ and $\rho_{i+1}$, $\Eb[X^{\pi'}|\Fc]$ can be any value lying in the lower and upper bounds. Since
\begin{align}
\Eb[X^{\pi_0}|\Fc]
=(i+d-1)\rho_1+\Eb[X^{\pi_0}|\Fc_{i+1,\tau_i+1}]\rho_2,
\end{align}
we can always find  $0\leq\rho_i,\rho_{i+1}\leq 1$ such that
\begin{align}\label{eqn:mean}
\Eb[X^{\pi'}|\Fc]=\Eb[X^{\pi_0}|\Fc].
\end{align}
Eqn.~(\ref{eqn:mean}) implies that
\begin{align}
\rho_{i+1}\rho_2(i+d) = &(\rho_{i+1}\rho_2-\rho_i\rho_1)\Eb[X^{\pi_0}|\Fc_{i+1,\tau_i+1}]\nonumber\\
&+ \rho_i\rho_1(i+d-1) \label{eqn:lemma4-1}.
\end{align}
Since $\Eb[X^{\pi_0}|\Fc_{i+1,\tau_i+1}]\geq \tau_i+d>i+d$, from Eqn.~(\ref{eqn:lemma4-1}), we must have
\begin{align}
\frac{\rho_i\rho_1}{\rho_{i+1}\rho_2}=\frac{\Eb[X^{\pi_0}|\Fc_{i+1,\tau_i+1}]-(i+d)}{\Eb[X^{\pi_0}|\Fc_{i+1,\tau_i+1}]-(i+d-1)}<1  .
\end{align}
Thus $\frac{\rho_{i+1}\rho_1-\rho_i\rho_1}{\rho_{i+1}\rho_2}$ and $\frac{\rho_i\rho_1}{\rho_{i+1}\rho_2}$ form a valid distribution. Based on Jensen's inequality, we have
\begin{align}
(i+d)^2<&\frac{\rho_{i+1}\rho_1-\rho_i\rho_1}{\rho_{i+1}\rho_2}(\Eb[X^{\pi_0}|\Fc_{i+1,\tau_i+1}])^2\nonumber\\
&+\frac{\rho_i\rho_1}{\rho_{i+1}\rho_2}(i+d-1)^2 \nonumber\\
\leq &\frac{\rho_{i+1}\rho_1-\rho_i\rho_1}{\rho_{i+1}\rho_2}\Eb[(X^{\pi_0})^2|\Fc_{i+1,\tau_i+1}]\nonumber\\
&+\frac{\rho_i\rho_1}{\rho_{i+1}\rho_2}(i+d-1)^2,
\end{align}
which implies that
\begin{align}
\rho_{i+1}\rho_2(i+d)^2 \leq & (\rho_{i+1}\rho_2-\rho_i\rho_1)\Eb[(X^{\pi_0})^2|\Fc_{i+1,\tau_i+1}]\nonumber\\
& + \rho_i \rho_1 (i+d-1)^2  .
\end{align}
Therefore,
\begin{align}\label{eqn:second_moment}
\Eb[(X^{\pi'})^2]\leq\Eb[(X^{\pi_0})^2].
\end{align}
Thus, $\pi_0$ cannot be optimal.

Following similar procedure, we can show that if $\tau_i=i$ or $\tau_i=i+1$, and the first update in that epoch arrives after slot $i$, the source will transmit the update until it is successfully delivered. 

Combining three cases, we conclude that the threshold $\tau_1$, $\tau_2\cdots$ are monotonically decreasing until $\tau_K=K$.

\section{Proof of Theorem~\ref{thm:non_uniform}}\label{appx:thm:non_uniform}
In the following, we prove the structural properties of the optimal policy characterized in Theorem~\ref{thm:non_uniform} one by one.
	
{\it Proof of Theorem~\ref{thm:non_uniform}(a).}  Assume at iteration $n$, the optimal action for state $\sv$ is to switch. It implies that $Q_{n-1}(\sv; 1)\leq Q_{n-1}(\sv; 0)$. Then, for state $\sv':=(\delta,l,c,b')$ with $b'<b$, we have $Q_{n-1}(\sv'; 1)\leq Q_{n-1}(\sv; 1)$, which is shown in the proof of Lemma~\ref{Lemma:nond}. Together with the fact that $Q_{n-1}(\sv'; 0)=Q_{n-1}(\sv; 0)$, the optimal action for state $\sv'$ at iteration $n$ is to switch as well. This property holds when $n\rightarrow\infty$, thus Theorem~\ref{thm:non_uniform}(a) is proved.
		
{\it Proof of Theorem~\ref{thm:non_uniform}(b).} We first consider the case when $b=l$. We have
\if{0}
If $b=l=1$, 
\begin{align*}
Q_n(\sv;1)&=\delta+\alpha\Eb[V^\alpha_n(1,0,0,\bar{b})],\\
Q_n(\sv;0)&=\delta+\alpha\Eb[V^\alpha_n(d,0,0,\bar{b})].
\end{align*} 
Therefore, $Q_n(\sv;1)\leq Q_n(\sv;0)$ according to Lemma~\ref{Lemma:nond}.
If $b=l>1$, 
\fi
\begin{align*}
Q_n(\sv;1)&=\delta+\alpha\Eb[V^\alpha_n(\delta+1,b-1,b,\bar{b})], \\
Q_n(\sv;0)&=\delta+\alpha\Eb[V^\alpha_n(\delta+1,l-1,c,\bar{b})].
\end{align*}
Based on the fact $l<c$ and the monotonicity of $V ^\alpha_n(\sv)$ in $c$ according to Lemma~\ref{Lemma:nond}, we must have $Q_n(\sv;1)\leq Q_n(\sv;0)$. 

Therefore, when $b=l$, the optimal action is to switch. Then, according to Theorem~\ref{thm:non_uniform}(a), when $b<l$, the optimal action is to switch as well.
	
{\it Proof of Theorem~\ref{thm:non_uniform}(c).} Theorem~\ref{thm:non_uniform}(c) can be proved in a way similar to the proof of Theorem~\ref{thm:non_uniform}(b), and is thus omitted.

{\it Proof of Theorem~\ref{thm:non_uniform}(d).}	
	To show the optimal decision is to switch, we need to prove $Q_k(\sv;1)\leq Q_k(\sv;0)$. 
\if{0}	
		Case 1: $b=1$. For this case,
	\begin{align}
	&\Delta Q_{n-1}^{1,0}(\delta,0,0,b)\nonumber\\
	&=\alpha\Eb[V^{\alpha}_{n-1}(1,0,0,\bar{b})]-\alpha\Eb[V^{\alpha}_{n-1}(\delta+1,0,0,\bar{b})],
	\end{align}
	which is non-positive according to Lemma~\ref{Lemma:nond}. 
	\fi
We note that
	\begin{align}\label{eqn:q01}
	Q_{n-1}((\delta,0,0,b);0)&=\delta+\alpha\Eb[V^{\alpha}_{n-1}(\delta+1,0,0,\bar{b})],\\
	Q_{n-1}((\delta,0,0,b);1)&=\delta+\alpha\Eb[V^{\alpha}_{n-1}(\delta+1,b\hspace{-0.03in}-\hspace{-0.03in}1,b,\bar{b})].\label{eqn:q11}
	\end{align}
	
	When $n< b$, initially, we have
	\begin{align} 
	\Eb[V^{\alpha}_0(\delta\hspace{-0.03in}+\hspace{-0.03in}n,b-n,b,\bar{b})]\hspace{-0.03in}\leq\hspace{-0.03in}\Eb[V^{\alpha}_0(\delta\hspace{-0.03in}+\hspace{-0.03in}n,0,0,\bar{b})].
	\end{align}
	Recursively applying Corollary~\ref{corollary:induction}, it gets
	\begin{align} 
	\Eb[V^{\alpha}_{n-1}(\delta\hspace{-0.01in}+\hspace{-0.01in}1,b-1,b,\bar{b})]\hspace{-0.02in}\leq\hspace{-0.02in}\Eb[V^{\alpha}_{n-1}(\delta+1,0,0,\bar{b})],
	\end{align}
	which indicates $\Delta Q_{n-1}^{1,0}(\sv)\leq 0$ according to (\ref{eqn:q01})-(\ref{eqn:q11}).
	
	When $n\geq b$, according to the monotonicity of $V(\sv)$ characterized in Lemma~\ref{Lemma:nond}, 
	\begin{align}
	\Eb[V^{\alpha}_{n-b}(b,0,0,\bar{b})]\leq\Eb[V^{\alpha}_{n-b}(\delta+b,0,0,\bar{b})].
	\end{align}
Then, based on the definition of $Q(\sv;0)$ in (\ref{eqn:w0}), we have
	\begin{align} 
	Q_{n\hspace{-0.02in}-\hspace{-0.02in}b}((\delta\hspace{-0.02in}+\hspace{-0.02in}b\hspace{-0.02in}-\hspace{-0.02in}1,1,b,.);0)\leq Q_{n\hspace{-0.02in}-\hspace{-0.02in}b}((\delta\hspace{-0.02in}+\hspace{-0.02in}b\hspace{-0.02in}-\hspace{-0.02in}1,0,0,.);0).
	\end{align}
Next, according to Lemma~\ref{Lemma:induction1}, 
		\begin{align} 
	\Eb[V^{\alpha}_{n\hspace{-0.02in}-\hspace{-0.02in}b\hspace{-0.02in}+\hspace{-0.02in}1}(\delta\hspace{-0.02in}+\hspace{-0.02in}b\hspace{-0.02in}-\hspace{-0.02in}1,1,b,\bar{b})]\hspace{-0.02in}\leq\hspace{-0.02in}\Eb[V^{\alpha}_{n\hspace{-0.02in}-\hspace{-0.02in}b\hspace{-0.02in}+\hspace{-0.02in}1}(\delta\hspace{-0.02in}+\hspace{-0.02in}b\hspace{-0.02in}-\hspace{-0.02in}1,0,0,\bar{b})].
		\end{align}
Applying Corollary~\ref{corollary:induction} recursively, 
	\begin{align} 
	\Eb[V^{\alpha}_{n-1}(\delta+1,b-1,b,\bar{b})] \leq\Eb[V^{\alpha}_{n-1}(\delta+1,0,0,\bar{b})].
	\end{align}
	which indicates $\Delta Q_{n-1}^{1,0}(\sv')\leq 0$ according to (\ref{eqn:q01})-(\ref{eqn:q11}).
	
	Thus, the optimal decision at state $(\delta,0,0,b)$ when $b\neq 0$ is always to switch.

{\it Proof of Theorem~\ref{thm:non_uniform}(e).}	 We first point out that if the optimal decision at state $\sv$ is to skip, it must have $b>l>0$ according to Theorem~\ref{thm:non_uniform}(b)(d). Thus, there are two possible cases depending on the value $l$:
	
Case 1: $l=1$. For this case,
	\begin{align}
	&\Delta Q_{n-1}^{1,0}(\delta,l,c,b)\nonumber\\
	&=\alpha\Eb[V^{\alpha}_{n-1}(\delta+1,b-1,b,\bar{b})]-\alpha\Eb[V^{\alpha}_{n-1}(c,0,0,\bar{b})],
	\end{align}
which is increasing in $\delta$ according to Lemma~\ref{Lemma:nond}. Thus, if $\Delta Q_{n-1}^{1,0}(\delta,l,c,b)\geq 0$, we must have $\Delta Q_{n-1}^{1,0}(\delta+1,l,c,b)\geq 0$.
	
Case 2: $l>1$. For this case, we have
	\begin{align}\label{eqn:q0}
	Q_{n-1}((\delta,l,c,b);0)&=\delta+\alpha\Eb[V^{\alpha}_{n-1}(\delta+1,l-1,c,\bar{b})],\\
	Q_{n-1}((\delta,l,c,b);1)&=\delta+\alpha\Eb[V^{\alpha}_{n-1}(\delta+1,b-1,b,\bar{b})].\label{eqn:q1}
	\end{align}
	
	When $n< l$, initially, we have
	\begin{align}
	Q_0((\delta\hspace{-0.035in}+\hspace{-0.035in}n,l\hspace{-0.03in}-\hspace{-0.03in}n\hspace{-0.03in}+\hspace{-0.03in}1,c,.);0)\hspace{-0.03in}\leq\hspace{-0.03in} Q_0((\delta\hspace{-0.03in}+\hspace{-0.03in}n,b\hspace{-0.03in}-\hspace{-0.03in}n\hspace{-0.03in}+\hspace{-0.03in}1,b,.);0).
	\end{align}
	Based on Lemma~\ref{Lemma:induction1}, we get
	\begin{align} 
	\Eb[V^{\alpha}_1(\delta\hspace{-0.03in}+\hspace{-0.03in}n,l\hspace{-0.03in}-\hspace{-0.03in}n\hspace{-0.03in}+\hspace{-0.03in}1,c,\bar{b})]\hspace{-0.03in}\leq\hspace{-0.03in}\Eb[V^{\alpha}_1(\delta\hspace{-0.03in}+\hspace{-0.03in}n,b\hspace{-0.03in}-\hspace{-0.03in}n\hspace{-0.03in}+\hspace{-0.03in}1,b,\bar{b})].
	\end{align}
Applying Corollary~\ref{Lemma:induction2}, we have
	\begin{align} 
	&\Eb[V^{\alpha}_2(\delta+n-1,l-n+2,c,\bar{b})]\nonumber\\
	&\leq \Eb[V^{\alpha}_2(\delta+n-1,b-n+2,b,\bar{b})].
	\end{align}
Recursively applying Corollary~\ref{Lemma:induction2}, it leads to
	\begin{align} 
	\Eb[V^{\alpha}_{n-1}(\delta\hspace{-0.01in}+\hspace{-0.01in}2,l-1,c,\bar{b})]\hspace{-0.02in}\leq\hspace{-0.02in}\Eb[V^{\alpha}_{n-1}(\delta+2,b-1,b,\bar{b})],
	\end{align}
	which indicates $\Delta Q_{n-1}^{1,0}(\sv')\geq 0$ according to (\ref{eqn:q0})-(\ref{eqn:q1}).
	
	When $n\geq l$, $\Delta Q_{n-1}^{1,0}(\sv)\geq 0$ implies that
	\begin{align}
	\Eb[V^{\alpha}_{n-1}(\delta\hspace{-0.01in}+\hspace{-0.01in}1,l\hspace{-0.01in}-\hspace{-0.01in}1,c,\bar{b})]\hspace{-0.01in}\leq\hspace{-0.01in}\Eb[V^{\alpha}_{n-1}(\delta\hspace{-0.01in}+\hspace{-0.01in}1,b-1,b,\bar{b})].
	\end{align}
according to (\ref{eqn:q0})-(\ref{eqn:q1}). 
Based on Corollary~\ref{Lemma:induction2}, we thus have
	\begin{align} 
	\Eb[V^{\alpha}_{n-2}(\delta+2,l\hspace{-0.01in}-\hspace{-0.01in}2,c,\bar{b})]\hspace{-0.01in}\leq\hspace{-0.01in}\Eb[V^{\alpha}_{n-2}(\delta\hspace{-0.01in}+\hspace{-0.01in}2,b\hspace{-0.01in}-\hspace{-0.01in}2,b,\bar{b})].
	\end{align}
We can further obtain the following inequality by applying Corollary~\ref{Lemma:induction2} recursively:
	\begin{align} 
	&\Eb[V^{\alpha}_{n-l+1}(\delta+l-1,1,c,\bar{b})]\nonumber\\
	& \leq\Eb[V^{\alpha}_{n-l+1}(\delta+l-1,b-l+1,b,\bar{b})].
	\end{align}
	Thus, according to Lemma~\ref{Lemma:induction1}, we have
	\begin{align} 
	&Q_{n-l}((\delta+l-1,1,c,.);0)\nonumber\\
	&\leq Q_{n-l}((\delta+l-1,b-l+1,b,.);0).
	\end{align}
Then, similar to the proof of Case 1, we have 
	\begin{align} 
	Q_{n-l}((\delta\hspace{-0.015in}+\hspace{-0.015in}l,1,c,.);0)\hspace{-0.015in}\leq\hspace{-0.015in} Q_{n-l}((\delta\hspace{-0.015in}+\hspace{-0.015in}l,b\hspace{-0.015in}-\hspace{-0.015in}l\hspace{-0.015in}+\hspace{-0.015in}1,b,.);0).
	\end{align}
Applying Lemma~\ref{Lemma:induction1} again, we have
	\begin{align} 
	\Eb[V^{\alpha}_{n-l+1}(\delta\hspace{-0.025in}+\hspace{-0.025in}l,1,c,\bar{b})]\hspace{-0.025in}\leq\hspace{-0.025in}\Eb[V^{\alpha}_{n-l+1}(\delta\hspace{-0.025in}+\hspace{-0.025in}l,b\hspace{-0.025in}-\hspace{-0.025in}l\hspace{-0.025in}+\hspace{-0.025in}1,b,\bar{b})].
	\end{align}
	We then apply Corollary~\ref{Lemma:induction2} recursively to obtain
	\begin{align} 
	\Eb[V^{\alpha}_{n-1}(\delta\hspace{-0.025in}+\hspace{-0.025in}2,l\hspace{-0.025in}-\hspace{-0.025in}1,c,\bar{b})]\hspace{-0.025in}\leq\hspace{-0.025in}\Eb[V^{\alpha}_{n-1}(\delta\hspace{-0.025in}+\hspace{-0.025in}2,b\hspace{-0.025in}-\hspace{-0.025in}1,b,\bar{b})],
	\end{align}
	which indicates $\Delta Q_{n-1}^{1,0}(\sv')\geq 0$ according to (\ref{eqn:q0})-(\ref{eqn:q1}).
	
	Combining both cases, we thus arrive at the conclusion that the optimal decision at state $\sv'$ is to skip as well.
	
\bibliographystyle{IEEEtran}
\bibliography{IEEEabrv,AgeInfo,ener_harv}
\epsfysize=3.2cm

\end{document}